\documentclass[a4paper,11pt]{article}
\usepackage{jheppub} 
\usepackage{lineno}
\usepackage{braket}
\usepackage{dsfont}
\usepackage{siunitx} 
\usepackage{mathtools}
\usepackage{float}
\usepackage{graphicx}

\preprint{TUM-EFT 196/25}

\title{\boldmath Thermalization of Bottomonium in the Quark--Gluon Plasma}

\author[a,b,c]{Nora Brambilla, }

\author[a]{Tom Magorsch}

\author[a]{and Antonio Vairo}

\affiliation[a]{Technical University of Munich, TUM School of Natural Sciences, Physics Department, James-Franck-Strasse 1, 85748 Garching, Germany}
\affiliation[b]{Institute for Advanced Study, Technische Universität München, Lichtenbergstrasse 2 a, 85748 Garching, Germany}
\affiliation[c]{Munich Data Science Institute, Technische Universität München, Walther-von-Dyck-Strasse 10, 85748 Garching, Germany}

\emailAdd{nora.brambilla@tum.de}
\emailAdd{tom.magorsch@tum.de}
\emailAdd{antonio.vairo@tum.de}

\abstract{
We study the approach to equilibrium of bottomonium in the quark-gluon plasma within the open quantum system framework. We perform large-scale simulations of the long-time behavior in three dimensions using the quantum trajectory method to observe the emergence of steady states and determine the timescale of thermalization in position-, angular-momentum-, and color-space. We find that the thermalization timescale increases with decreasing temperature and decreasing coupling to the medium, which is given by transport coefficients of the medium. Additionally, we observe that the steady states exhibit small corrections to the Gibbs state due to medium interactions and show that these corrections diminish for weaker medium coupling and higher temperature. At a temperature of $\SI{450}{\MeV}$, quarkonium relaxes to a state that is approximately thermal, with the most significant correction being a smaller overlap of the $1S$ state with respect to the Gibbs state. We compare these findings with the master equation obtained at leading order in the expansion of the binding energy over the temperature, which we find to have a trivial steady state.
}

\begin{document}
\maketitle
\flushbottom

\section{Introduction}

Quarkonium has long been an essential probe of the quark-gluon plasma (QGP) created in heavy-ion collisions.
The interaction between quarkonium and the QGP changes the dynamics of quarkonium while it traverses the plasma, leaving its imprints on the specifics of the evolution.
A detailed investigation of the quarkonium dynamics, therefore, enables us to draw conclusions about the QGP.
In particular, the phenomenon of quarkonium suppression was proposed by Matsui and Satz as an experimental signature to probe the existence of the QGP~\cite{Matsui:1986dk}.
In their original work, thermal screening of the quarkonium potential would lead to the quarkonium states becoming unbound for increasing temperatures, modifying the number of measured quarkonia in heavy-ion collision experiments.
Within the last decades, this static picture has been superseded by a more dynamical view of the evolution of quarkonium in a thermal medium~\cite{Rothkopf:2019ipj}.
Opposed to the static picture, the melting of quarkonium is a time-dependent process that is not in equilibrium. Understanding the dynamics of quarkonium within the deconfined medium and the approach to equilibrium is a key problem in quarkonium studies.
In recent years, a novel paradigm has emerged to describe this dynamical process. Using non-relativistic effective field theories in combination with the open quantum system framework allows to derive quantum master equations for the real-time evolution of the quarkonium~\cite{Akamatsu:2020ypb,Yao:2021lus,Andronic:2024oxz}.
Different master equations have been derived from non-relativistic QCD (NRQCD)~\cite{Bodwin:1994jh} and potential non-relativistic QCD (pNRQCD)~\cite{Brambilla:1999xf,Brambilla:2004jw}, and numerical simulations were able to describe bottomonium observables in heavy-ion collision experiments~\cite{Miura:2019ssi,Blaizot:2017ypk,Sharma:2019xum,Brambilla:2023hkw,Brambilla:2024tqg,Yao:2018nmy,Yao:2020eqy}.
While the framework is capable of making phenomenological predictions, it also enables a deeper insight into the interplay between the properties of the medium and the relaxation dynamics of the quarkonium. 

One key question concerns the thermalization of the heavy probes inside the medium. While measurements of the elliptic flow suggest the thermalization of charm quarks, so far, bottomonium does not show a significant elliptic flow, suggesting that bottomonium does not fully equilibrate within the lifetime of the QGP~\cite{Greco:2003vf,ALICE:2017pbx,CMS:2017vhp,ALICE:2019pox}. 
While due to the larger abundance of charm quarks, the recombination of free charm plays an important role in the chemical equilibration of charmonium, in the bottomonium case, the particles are sufficiently dilute, with interactions occurring rarely between different bound states. In this dilute limit, regeneration of uncorrelated bottom quarks is negligible, which also means dissociation effects dominate within heavy-ion collisions.
Nevertheless, a single bottomonium particle will still exchange energy with the QGP, thus driving internal degrees of freedom towards thermal equilibrium, while its center-of-mass momentum simultaneously relaxes towards a state of kinetic equilibrium with the plasma.

This work aims to study the bottomonium's thermalization process in a thermal medium within the open quantum system framework. 
Previous studies investigated the open Schwinger model, observing string breaking and the thermalization to a steady state close to the thermal state~\cite{deJong:2021wsd,Lee:2023urk,Lin:2024eiz,Angelides:2025hjt}.
Furthermore, a simulation of a master equation derived from NRQCD observed the equilibration of bottomonium to the Boltzmann distribution within \SI{150}{\femto\meter}~\cite{Miura:2022arv}. In contrast, the solution of a quantum master equation for charmonium revealed an approach to an equilibrium steady state with an offset from the Boltzmann distribution~\cite{Delorme:2024rdo}.
However, these studies were carried out in a single spatial dimension, therefore neglecting the dynamics of the angular momentum sector. In this paper, we provide a detailed study of the approach to the equilibrium of bottomonium in a thermal medium under the open quantum system paradigm. We focus on master equations of the Lindblad type, obtained from pNRQCD in three dimensions~\cite{Brambilla:2016wgg,Brambilla:2017zei,Brambilla:2022ynh}, and deploy large-scale simulations to answer the following questions:
\begin{enumerate}
    \item Is there an equilibrium state?
    \item At what timescales is it reached?
    \item How does the steady state look like?
\end{enumerate}
In particular, the simuliations are performed with sufficieintly small lattice spacings to ensure that discretization effects remain small. To our knowledge, open quantum system simulations of quarkonium in medium with such small lattice spacings and long evolution times have not been performed before.

The paper is organized as follows: In section~\ref{sec:theory}, we present the theoretical background for our study by introducing the Lindblad equation for the bottomonium evolution in section~\ref{sec:mastereqs} and recalling the Liouvillian formalism for the Lindblad equation in section~\ref{sec:sstates}. In section~\ref{sec:proofs}, we discuss the existence and uniqueness of an attractive steady state. In section~\ref{sec:timescale}, we present simulation results for the time-evolution, identify equilibration timescales, and investigate the dependence on different simulation parameters. In section~\ref{sec:steady_state}, we inspect the structure of the steady state and compare it with the Gibbs state.
Finally, in section~\ref{sec:conc}, we summarize and conclude.
Furthermore, we clarify some technical details in the appendices. In appendix~\ref{app:proofs}, we elaborate on the uniqueness of the steady state by checking the uniqueness condition under some special conditions. In appendix~\ref{app:dt}, we investigate the dependence of the simulation on the time discretization and compare with simulations subject to a different timescale and temperature evolution more relevant to the phenomenology of quarkonium suppression in heavy-ion collisions. Finally, in appendix~\ref{app:1d}, we comment on one-dimensional master equations.

\section{Theory}
\label{sec:theory}

\subsection{Quarkonium in the QGP as an open quantum system}
\label{sec:mastereqs}
Quarkonium suppression in a medium can be studied systematically by utilizing non-relativistic effective field theories to exploit the hierarchy of scales present in the system. Specifically, the combination of thermal field theory with the open quantum system framework allows one to derive quantum master equations for the quarkonium density matrix. In this work, we focus on the evolution equations obtained from pNRQCD in~\cite{Brambilla:2016wgg,Brambilla:2017zei,Brambilla:2022ynh}. The resulting evolution equations are applicable under the hierarchy of scales $M\gg 1/a_0 \gg \pi T$, where $M$ is the heavy quark mass, $a_0$ is the Bohr radius of the quarkonium state in consideration, and $T$ is the temperature of the medium. Furthermore, to bring the evolution equations into Lindblad form~\cite{Lindblad:1975ef,Gorini:1975nb}, the quantum Brownian regime, characterized by $\pi T \gg E$, with $E$ the binding energy, is utilized. 
In particular, the hierarchy $1/a_0 \gg \pi T \sim m_D \gg E $,
 where $m_D$ is the Debye mass well describes bottomonia states with a small radius evolving in a strongly coupled QGP.
The Lindbladian evolution is completely positive, guaranteeing the density matrix will be physical throughout evolution. In the following, we consider a Lindblad equation for the quarkonium density matrix $\rho$ of the form 
\begin{equation}
    \frac{d\rho(t)}{dt}=\mathcal{L}\left[\rho(t)\right]=-i[H,\rho(t)]+\sum_{n,i}\left[C_i^n\rho(t)C_i^{n\dagger}-\frac{1}{2}\left\{C_i^{n\dagger}C_i^n,\rho(t)\right\}\right],
    \label{eq:lindblad}
\end{equation}
where $H$ is the Hamiltonian and the $C^n_i$ denote the Lindblad operators associated with the spatial dimensions $i=x,y,z$. The three-dimensional Lindblad equation for the evolution of bottomonium in a strongly coupled thermal medium, as derived from pNRQCD at next-to-leading order in the binding-energy over temperature~\cite{Brambilla:2022ynh}, is given by the Hamiltonian
\begin{equation}
H=
\begin{pmatrix} 
h_s+ \displaystyle\frac{r^2}{2}\gamma +\frac{\kappa}{4MT}\{r_i,p_i\} & 0 \\
0 & \displaystyle h_o+\frac{N^2_c-2}{2(N^2_c-1)}\left(\frac{r^2}{2}\gamma + \frac{\kappa}{4MT}\{r_i,p_i\}\right)
\end{pmatrix},
\label{eq:HamiltonianQCDNLO}
\end{equation}
and the Lindblad operators
\begin{align}
    C_i^0 &= \sqrt{\frac{\kappa}{N^2_c-1}}\begin{pmatrix}
        0 & r_i+\frac{ip_i}{2MT}+\frac{\Delta V_{os}}{4T}r_i\\
        \sqrt{N^2_c-1}\left(r_i+\frac{ip_i}{2MT}+\frac{\Delta V_{so}}{4T}r_i\right) & 0
    \end{pmatrix},\label{eq:QCDNLOC0}\\
    C_i^1 &= \sqrt{\frac{\kappa(N^2_c - 4)}{2(N^2_c-1)}}\begin{pmatrix}
        0 & 0\\
        0 & r_i+\frac{ip_i}{2MT}\end{pmatrix},\label{eq:QCDNLOC1}
\end{align}
where $r^2=\sum_ir^2_i$ and $p_i$ are the momentum components. The vacuum Hamiltonians are defined as $h_{s(o)}=p^2/M+V_{s(o)}(r)$, with $V_{s(o)}(r)$ the color singlet and the color octet potentials. At leading order, these potentials are an attractive ($-(N^2_c-1)\alpha_s/(2N_cr)$) and repulsive ($\alpha_s/(2N_cr)$) Coulomb potential, respectively, which we will use for our study. We also define $\Delta V_{uw} = V_u(r) - V_w(r)$. Several quarkonium transport coefficients may couple the quarkonium with the medium (see e.g. ref.~\cite{Scheihing-Hitschfeld:2023tuz}). Within some approximations~\cite{Brambilla:2024quarkonium}, we assume that the coupling with the medium may be reduced to only two coefficients $\kappa$ and $\gamma$, which are the heavy quarkonium momentum diffusion coefficient and its dispersive counterpart, respectively. The $2\times 2$ matrices reflect the singlet and octet structure in color space. Finally, $N_c$ is the number of colors, which, for QCD, is set to $N_c=3$. To study the effect of higher order terms in the $E/(\pi T)$ expansion, we furthermore consider the master equation at leading order.

\paragraph{pNRQCD at LO in $E/(\pi T)$.}
At leading order in the binding energy over temperature expansion, we drop the terms suppressed by $E/(\pi T)$ in the power counting, resulting in a Lindblad equation with
\begin{equation}
H=
\begin{pmatrix} 
h_s+ \displaystyle\frac{r^2}{2}\gamma & 0 \\
0 & \displaystyle h_o+\frac{N^2_c-2}{2(N^2_c-1)}\frac{r^2}{2}\gamma
\end{pmatrix},
\label{eq:HamiltonianLO}
\end{equation}
and the Lindblad operators
\begin{align}
    C_i^0 &= \sqrt{\frac{\kappa}{N^2_c-1}}r_i\begin{pmatrix}
        0 & 1\\
        \sqrt{N^2_c-1} & 0
    \end{pmatrix},\label{eq:C0QCDLO}\\
    C_i^1 &= \sqrt{\frac{\kappa(N^2_c - 4)}{2(N^2_c-1)}}r_i\begin{pmatrix}
        0 & 0\\
        0 & 1
    \end{pmatrix}.
    \label{eq:C1QCDLO}
\end{align}

\vspace{1em}
\noindent For both master equations, it is possible to project the three-dimensional equation onto spherical harmonics, see, e.g., ref.~\cite{Brambilla:2022ynh}. Assuming isotropy, the resulting equation describes the evolution of the density matrix in a single spatial dimension, given by the radial coordinate $r$ and the angular momentum quantum number $l$. The Lindblad operators induce transitions between different angular momentum states and transitions between singlet and octet states.

\subsection{Steady states of the Lindblad equation}
\label{sec:sstates}

The problem of the long-time behavior for the Lindblad equation can be understood by considering the equation under the Choi--Jamiolkowski isomorphism~\cite{Choi:1975nug,Jamiolkowski:1972pzh}. This amounts to writing the density matrix $\rho$ as
\begin{equation}
  \label{eq:4}
  \rho = \sum_{i,j}\ket{i}\bra{j} \to\text{vec}(\rho) = \sum_{i,j}\ket{j}\otimes\ket{i}.
\end{equation}
In the following, we denote the vectorized density matrix as $\hat\rho\coloneqq\text{vec}(\rho)$ and, in general, the vectorized version of any other operator $O$ similarly as $\hat{O}\coloneqq\text{vec}(O)$. The Lindblad equation~\eqref{eq:lindblad} then takes the form 
\begin{equation}
    \frac{d}{dt}\hat\rho = \hat{\mathcal{L}}\hat\rho,
    \label{eq:lindblad2}
\end{equation}
with the Liouvillian superoperator $\hat{\mathcal{L}}$ defined as
\begin{equation}
  \label{eq:liouvillian}
  \hat{\mathcal{L}} = -i(I\otimes H - H^{T}\otimes I) + \sum_{n,i}\left[C^{n*}_{i}\otimes C^n_{i} - \frac{1}{2}I\otimes C^{n\dagger}_{i}C^n_{i} - \frac{1}{2}(C^{n\dagger}_{i}C^n_{i})^{T}\otimes I\right],
\end{equation}
where $I$ denotes the identity. Figuratively speaking, for a finite Hilbert space, this corresponds to stacking columns of the density matrix to obtain a vector of size $N^2_H$, where $N_H$ is the size of the Hilbert space. The Liouvillian $\hat{\mathcal{L}}$ then corresponds to a $N^2_H\times N^2_H$ matrix.
The time evolution of the density matrix is formally given by the solution of eq.~\eqref{eq:lindblad2} as
\begin{equation}
    \hat\rho(t)=e^{\hat{\mathcal{L}}t}\hat\rho(0).
    \label{eq:LiouTimeEvolve}
\end{equation}

We can now deduce some characteristics of the time evolution from the spectrum of the Liouvillian $\hat{\mathcal{L}}$. Since $\hat{\mathcal{L}}$ is not Hermitian it has separate right and left eigenvectors $x_a$ and $y_a$ with eigenvalues $\lambda_a$~\cite{Brody:2013axr} as
\begin{align}
  \hat{\mathcal{L}}x_{a} &= \lambda_{a} x_{a},\\
  y^{\dagger}_{a}\hat{\mathcal{L}} &= \lambda_{a} y^{\dagger}_{a}.
\end{align}
Here, we have used that there are always two left and right eigenstates with the same eigenvalues $\lambda_a$, hence we can relabel them with the same index. 
If the Liouvillian is diagonalizable, we can then express it as 
\begin{equation}
  \label{eq:10}
  \hat{\mathcal{L}} = \sum_{a}\lambda_{a}x_{a}y^{\dagger}_{a},
\end{equation}
so that
\begin{equation}
  \label{eq:spectralTimeEvolve}  e^{\hat{\mathcal{L}}t}=\sum_{a}e^{\lambda_{a}t}x_{a}y^{\dagger}_{a},
\end{equation}
dictates the time evolution. We know that every zero eigenvalue $\lambda_a=0$ has a corresponding steady state solution to eq.~\eqref{eq:LiouTimeEvolve} given by $x_a$ since $\hat{\mathcal{L}}x_a = 0$. 
It is, therefore, possible to numerically obtain the exact steady-state solutions of a Lindblad equation by diagonalizing the $N^2_H\times N^2_H$ Liouvillian~\eqref{eq:liouvillian}. However, this is only practically achievable for sufficiently small Hilbert spaces.
One can prove that there is always at least one steady state, which follows from the trace-preserving nature of the Lindblad equation, by noticing that
\begin{equation}
    \frac{d}{dt}\text{Tr}(\rho(t))=\frac{d}{dt}\hat{I}^\dagger\hat\rho(t)= \hat{I}^\dagger\hat{\mathcal{L}}\hat\rho(t)=0,
\end{equation}
where $\hat{I}$ denotes the identity in the Choi--Jamiolkowski isomorphism.
Since this equation has to be fulfilled for every $\hat\rho(t)$ we know that 
\begin{equation}
\hat{I}^\dagger\hat{\mathcal{L}}=0.
\end{equation}
The property of trace preservation, therefore, leads to the identity $\hat{I}$ always being a left eigenvector of $\hat{\mathcal{L}}$ with eigenvalue zero. The corresponding right eigenvector $x_I$ will also have the eigenvalue zero, leading to 
\begin{equation}
    \hat{\mathcal{L}}x_I = 0,
\end{equation}
which means $dx_I/dt=0$ and $x_I$ can be identified as a steady state.
Using the spectral decomposition~\eqref{eq:spectralTimeEvolve}, we can further understand the anatomy of the time evolution. For a given initial state, the time evolution becomes
\begin{equation}
    \hat{\rho}(t)=\sum_a c_ae^{\lambda_at}x_a,
    \label{eq:spectralrhot}
\end{equation}
where the coefficients $c_a$ depend on the initial state $c_a=y^\dagger_a\hat{\rho}(0)$. Equation~\eqref{eq:spectralrhot} implies that $\text{Re}(\lambda_a)\leq 0$ since positive eigenvalues would lead to unphysical exponential growth. If $\text{Re}(\lambda_a)<0$, the contribution of $x_a$ to the time evolution dies off in the large time limit, and only contributions with eigenvalue zero can survive. From this, we conclude that if there is only a single unique steady state, the Lindblad equation will necessarily relax towards it. We note that, in principle, purely imaginary eigenvalues would lead to circular trajectories of the density matrix. However, it has been shown that such contributions are not possible~\cite{Baumgartner_2008_I}.
If there are multiple steady states, the space of density matrices is divided into subspaces, each of which relaxes to a different steady state~\cite{Baumgartner_2008}. 

Given the spectrum of the Liouvillian, eq.~\eqref{eq:spectralrhot} also provides a measure for the timescale of thermalization. Since the non-zero eigenvalue with the lowest absolute real part corresponds to the exponential $\exp(\lambda_a t)$, which decays the slowest, it sets the timescale needed to reach the steady state. This difference between the zero and the second-largest eigenvalue is also called the Liouvillian gap. While in the case of a finite-dimensional Hilbert space, $\hat{\mathcal{L}}$ has a discrete spectrum and, therefore, a finite Liouvillian gap, in the case of an infinite Hilbert space, $\hat{\mathcal{L}}$ can in principle have a continuous spectrum including eigenvalues that are arbitrarily close to zero. In these cases, the Liouvillian gap closes, and, while there can still be a unique, attractive state, it can happen that the approach to it is slower than exponential.

A priori, it is not clear if a given Lindblad equation has one or more steady states; however, there have been different conditions developed to check the uniqueness of the steady state of a Lindblad equation~\cite{Nigro:2018xov,Spohn_1977,Frigerio:1978gu}.
In particular, we consider the following necessary and sufficient condition: A Lindblad equation with Hamiltonian $H$ and Lindblad operators $C_n$ has a unique steady state iff the commutant $\{C_n,C^\dagger_n,H\}^\prime$, which is the set of all operators commuting with all operators in $\{C_n,C^\dagger_n,H\}$ is trivial, i.e. it contains operators proportional to the identity~\cite{Spohn_1977,Frigerio:1978gu,Evans:1977jg}.
An operator commuting with all operators in the set would generate an invariant subspace under time evolution, containing a different steady state.
Using this powerful condition, we can investigate if a Lindblad equation has a unique steady state, and by the spectral decomposition, we know that, in this case, the dynamics would necessarily relax towards it. 

At this point, while we can check if the system is equilibrating towards a unique steady state, we do not know what this steady state looks like. Under the open quantum system paradigm, one might assume that the composite system-bath compound is embedded in a super-bath, which keeps the plasma at temperature $T$~\cite{Trushechkin:2021drp,Purkayastha:2024paj}. In this case, one would expect that the steady state of the system is the Gibbs state of the system-environment Hamiltonian $H_{SE}$, traced over the environmental degrees of freedom
\begin{equation}
\rho=\text{Tr}_E\left(\frac{1}{Z}e^{-H_{SE}/T}\right).
\end{equation}
In the case of a vanishing system-environment interaction, this reduces to the expected result, which is the Gibbs state for the system
\begin{equation}
    \rho_\text{Gibbs}=\frac{1}{Z}e^{-H_S/T},
    \label{eq:Gibbs}
\end{equation}
where $H_S$ is the system Hamiltonian. In the master equation presented in eqs.~\eqref{eq:lindblad}-\eqref{eq:QCDNLOC1}, $H_S$ would correspond to the vacuum Hamiltonian $h_s$. For finite interactions, however, the corrections to the Gibbs state of the system can be sizable.

\section{Results}
\label{sec:results}

\subsection{Is there an equilibrium state?}
\label{sec:proofs}
In section~\ref{sec:sstates}, we established that the steady states of a Lindblad equation are attractive, and thus, a Lindblad equation with a unique steady state will necessarily relax to this state in the long-time limit. The condition for uniqueness~\cite{Spohn_1977,Frigerio:1978gu,Evans:1977jg} presented in section~\ref{sec:sstates} is that there is no operator $O$ other than $O\propto\mathds{1}$ that fulfills all the conditions
\begin{align}
    [H,O]&=0, \label{eq:cond1} \\ 
    [C_i,O]&=0,\label{eq:cond2} \\ 
    [C^\dagger_i,O]&=0. \label{eq:cond3}
\end{align}
Because if $O$ satisfies \eqref{eq:cond1}-\eqref{eq:cond3} also $O^\dagger$ does, we can restrict to $O$ being Hermitian. Hermitian operators that commute with $H$ identify conserved observables with respect to the Hamiltonian $H$. The only conserved observables in a central potential, which corresponds to $H$ at LO in $E/(\pi T)$, are the angular momentum operators
\begin{equation}
    L_i = \epsilon_{ijk}r_jp_k.
\end{equation}
In the special case of the Coulomb potential, there are additionally the components of the Laplace–Runge–Lenz vector
\begin{equation}
    A_i = \frac{1}{2M}\epsilon_{ijk}(
    p_jL_k-L_jp_k)-\alpha\frac{r_i}{r},
\end{equation}
where $\alpha=-(N^2_c-1)\alpha_s/(2N_c)$ is the Coulombic coupling constant. 
Considering Lindblad operators of the form given in eqs.~\eqref{eq:QCDNLOC0} and~\eqref{eq:QCDNLOC1}, it is clear that neither the Hamiltonian, nor the angular momentum, nor the Laplace--Runge--Lenz vector components commute with all Lindblad operators simultaneously. Lastly, the parity operator also commutes with the Hamiltonian, however, it does not commute with the Lindblad operators. This suggests the view that, for a Coulombic potential, a master equation with Lindblad operators of this form leads to a unique steady state. We elaborate further on this in appendix~\ref{app:proofs}. If there is only a single zero eigenvalue in the spectrum of the Liouvillian, this guarantees the evolution towards the corresponding eigenvector. It implies the loss of memory of the system's initial conditions, which is a characteristic feature of thermalization and ensures that the steady state results are independent of the chosen initial state.

\subsection{At what timescales does equilibration happen?}
\label{sec:timescale}

To study the approach to equilibrium, we simulate the evolution of bottomonium inside a plasma in three dimensions as dictated by the Lindblad equation~\eqref{eq:lindblad}. The simulation is performed using the open source code QTraj~\cite{Omar:2021kra}, based on the quantum trajectories algorithm~\cite{Molmer:1993ltv}. In our simulations, we use a box of size $L=20\text{GeV}^{-1}$ and $N=1024$ lattice points corresponding to a lattice spacing of $dr=\SI{0.004}{\femto\meter}$. For the transport coefficient, we choose the values $\kappa=4\,T^3$ and $\gamma=0$, consistent with bottomonium suppression data~\cite{Brambilla:2023hkw}. The initial state is chosen as a singlet with angular momentum $l=0$ and a localized Gaussian peak in position space. Furthermore, for the strong coupling we choose $\alpha_s=0.468$ based on previous studies~\cite{Brambilla:2022ynh}. For more details on the computational setup, we refer the reader to refs.~\cite{Brambilla:2022ynh,Omar:2021kra}. 
Here, we note that in order to obtain accurate results for long time frames, the time-step $\delta t$ needs to be chosen sufficiently small. For too large $\delta t$, especially the accuracy of the high angular momentum states will suffer, which can be traced back to the fact that higher angular momentum states have a large width $\Gamma=\sum C_i^{n\dagger} C^n_i$. A large width leads to two effects: Firstly, in the quantum trajectories framework, quantum jumps are triggered a lot faster, and a smaller $\delta t$ is required to resolve them in order not to underestimate the number of jumps. Secondly, to evolve the wave functions, an implicit expansion in $\delta tH_\text{eff}$ with $H_\text{eff}=H-i\Gamma/2$ is performed, whose convergence worsens for large $\Gamma$, rendering the simulation less accurate for larger $l$. 
In practice, to conserve computational resources, we impose a maximum angular momentum $l_\text{max}$ and set a boundary condition, that trajectories with $l=l_\text{max}$ transition to $l_\text{max}-1$ with probability $p^\downarrow_l$ and stay at $l=l_\text{max}$ with probability $p^\uparrow_l$, where $p^{\uparrow\downarrow}_l$ are the transition probabilities in the quantum trajectory framework.
The truncation to a finite angular momentum space is similar to restricting to a finite box in position space.
In phenomenological studies, the relevant observables, specifically the $S$-wave overlaps, are less dependent on $\delta t$, as, e.g., shown in the appendix of ref.~\cite{Omar:2021kra}. We further study the dependence on $\delta t$ in appendix~\ref{app:dt}.

To assess the non-equilibrium evolution of the full-density matrix, we consider four different observables. Firstly, we compute the overlaps with quarkonium eigenstates $\ket{r,nS}$,
\begin{equation}
    O_n = \int dr \int dr^\prime \braket{r,nS|\rho(t)|r^\prime,nS},
\end{equation}
for the lowest $S$-wave quarkonium states $n=1,2,3$. These overlaps are directly proportional to the quarkonium survival probabilities relevant to quarkonium suppression. In addition, we compute the occupations of different angular momentum and color states, $\ket{r,l,c}$, given by 
\begin{equation}
    P_l(t) = \sum_c\int dr \braket{r,l,c|\rho(t)|r,l,c},
\end{equation}
and 
\begin{equation}
    P_c(t) = \sum_l\int dr\braket{r,l,c|\rho(t)|r,l,c},
\end{equation}
respectively. In the quantum trajectories framework, these are given by the fraction of trajectories with angular momentum $l$ or color $c$ at a given timestep.

For all these observables, we can make a prediction assuming that the system is in the Gibbs state $\rho_\text{Gibbs}$ as
\begin{align}
    O_n &= \frac{Z_{n0s}}{Z_\text{tot}},\label{eq:Oi_gibs}\\
    P_l &= \frac{\sum_{nc}Z_{nlc}}{Z_\text{tot}},\label{eq:Pl_gibs}\\
    P_c &= \frac{\sum_{nl}Z_{nlc}}{Z_\text{tot}},\label{eq:Pc_gibs}
\end{align}
where for each energy quantum number $n$, angular momentum $l$ and color $c$, we define in the color singlet case ($c=s$) $Z_{nls}=(2l+1)\exp(-E_{nls}/T)$ and in the color octet case $(c=o)$ $Z_{nlo}=8(2l+1)\exp(-E_{nlo}/T)$, where the factor in front of the exponential accounts for the degeneracy of the energies in the magnetic and color quantum numbers. The partition function is then $Z_\text{tot}=\sum_{nlc}Z_{nlc}$. The quantity $E_{nlc}$ denotes the energy, which depends on $n$, $l$ and $c$. The dependence on $c$ arises since, for the singlet and the octet, we have different potentials leading to different energies. We then numerically solve the Schrödinger equation
\begin{equation}
    h_{lc}\ket{\psi_n} = E_{nlc}\ket{\psi_n},
\end{equation}
where 
\begin{equation}
    h_{lc} = \frac{p^2}{M}+V_c(r)+\frac{l(l+1)}{Mr^2},
\end{equation}
is the in-vacuum Hamiltonian for a given angular momentum and color state. Here $p^2$ is the radial part of the square of the momentum operator, which, in the spherical basis, can be written as $p^2=-(\partial^2/\partial r^2 + 2/r\, \partial/\partial r)$. Without an angular momentum cutoff $l_\text{max}$ or a finite discretized simulation volume, the partition sums would diverge. With suitable cutoffs, however, we obtain finite predictions for all observables.

Lastly, we examine the position space density matrix for a given angular momentum and color quantum number by projecting on the states $\ket{r,l,c}$
\begin{equation}
    \rho^{l}_{c}(r,r^\prime)=\braket{r,l,c|\rho(t)|r^\prime,l,c}.
\end{equation}
In practice, we obtain $\rho^{l}_{c}(r,r^\prime)$ from our simulations by saving to disk the position space wave function $\psi^l_{c,k}(r)$ and color and angular momentum quantum numbers $c$ and $l$ for each quantum trajectory at specific times. 
We then obtain the density matrix as
\begin{equation}
    \rho^{l}_{c}(r,r^\prime)=\frac{1}{N}\sum^N_k \psi^l_{c,k}(r)\psi^{l\dagger}_{c,k}(r^\prime),
\end{equation}
where $k$ labels the individual trajectories and $N$ is the total number of trajectories sampled with the given quantum numbers.

Finally, we comment on the angular momentum behaviour in the three-dimensional case. When a jump is triggered, at leading order in $E/(\pi T)$ transition probabilities for going up or down in angular momentum are given by~\cite{Brambilla:2017zei}
\begin{align}
    p^{\uparrow}_l = \frac{l+1}{2l+1},\label{eq:lopup}\\
    p^{\downarrow}_l = \frac{l}{2l+1},\label{eq:lopdown}
\end{align}
respectively. Since $p^{\uparrow}_l>p^{\downarrow}_l$ for all $l$, the average angular momentum will continuously increase, and for any cutoff $l_\text{max}$, the steady state distribution of $P_l$ will increase monotonically in $l$. This disagrees with the expectation from the Gibbs state~\eqref{eq:Pl_gibs}, where we expect $P_l\to 0$ for $l\to\infty$. At next-to-leading order in $E/(\pi T)$, the probabilities $p^{\uparrow\downarrow}_l$ are not given by Eqs.~\eqref{eq:lopup} and~\eqref{eq:lopdown} but rather they take the form\footnote{These formulas correct those present in the original version of~\cite{Brambilla:2022ynh}. We thank Arthur Lin for detecting the error.}
\begin{align}
    &p^{\uparrow}_l = \frac{l+1}{2l+1}\frac{\braket{A(l)}-(l+1)\braket{B}}{\braket{A(l)}-\braket{B}},\label{eq:nlopup}\\
    &p^{\downarrow}_l = \frac{l}{2l+1}\frac{\braket{A(l)}+l\braket{B}}{\braket{A(l)}-\braket{B}},\label{eq:nlopdown}
\end{align}
where $\braket{A(l)}$ and $\braket{B}$ are expectations on the state of some operators $A$ and $B$, which can be derived from the width.
The operator $A$ explicitly depends on the angular momentum, while $B$ is independent of $l$. Since the expectations depend on the specific position space evolution of the state, evaluating these probabilities leads to non-trivial results. Due to the positivity of the width, it must hold that
\begin{equation}
\braket{A(l)}\geq(l+1)\braket{B}.
\end{equation}
For the probabilities~\eqref{eq:nlopup} and~\eqref{eq:nlopdown} differently from~\eqref{eq:lopup} and~\eqref{eq:lopdown}, and depending on the exact scaling of $\braket{A(l)}$ and $\braket{B}$, it is possible that $p^\downarrow_l\geq p^\uparrow_l$ for large $l$. In the next-to-leading order formulation, therefore, a more consistent behaviour of the angular momentum sector is possible. We analyze the quantitative agreement of $P_l$ between the Gibbs state and the next-to-leading order steady state in section~\ref{sec:steady_state}.

\paragraph{Approach to equilibrium of bottomonium.}

To examine the approach to equilibrium of bottomonium in a thermal medium, we simulate the pNRQCD master equation at NLO in $E/(\pi T)$ presented in section~\ref{sec:mastereqs} at a constant temperature of $T=\SI{450}{\MeV}$. For all cases, we sample approximately $\num{3e5}$ trajectories to ensure sufficiently small statistical noise, allowing us to reliably observe the emergence of a steady state. To assess the dependence on the angular momentum cutoff, we perform separate simulations with $l_\text{max}=3,6,9$.

\begin{figure}[h]
    \centering
    \includegraphics[width=0.95\textwidth]{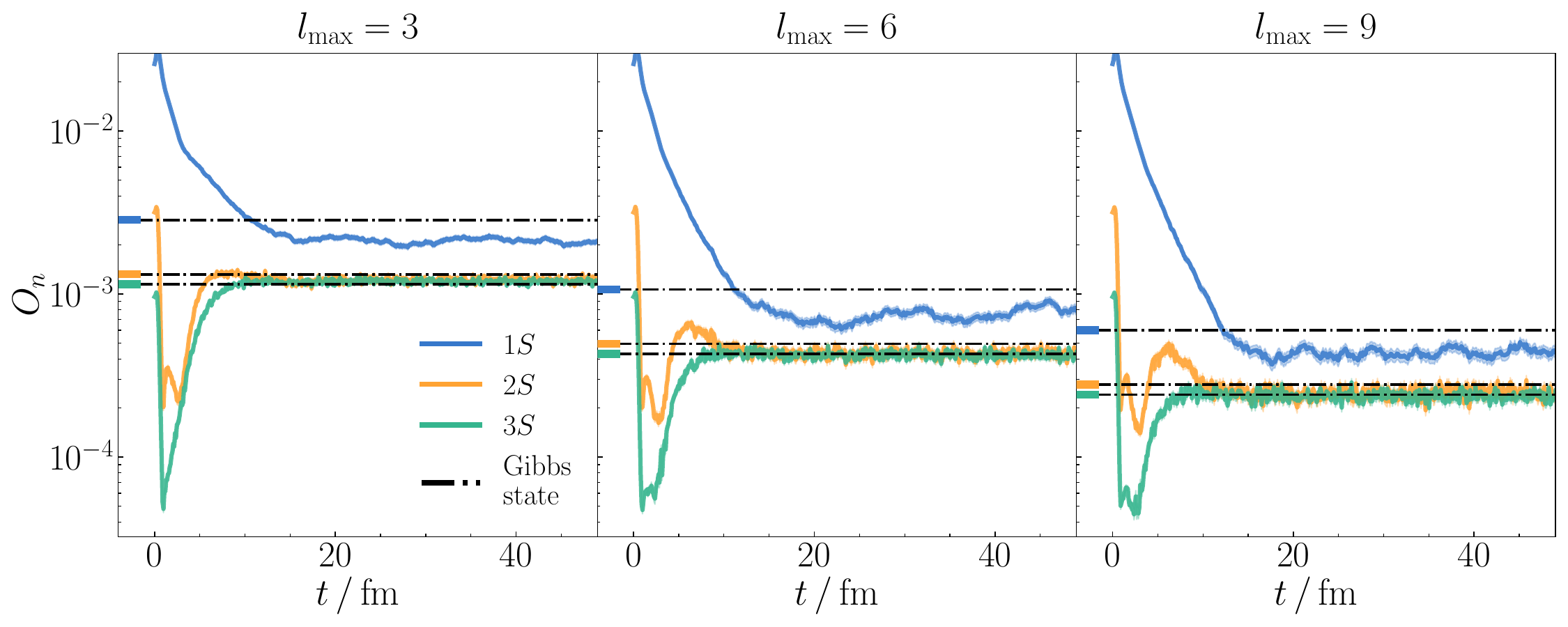}
    \caption{The time evolution of $S$-wave overlaps $O_n$ with $n=1S$ {(blue)}, $2S$ {(orange)} and $3S$ {(green)} quarkonium eigenfunctions from the simulation of the pNRQCD master equation at NLO in $E/(\pi T)$ at a constant temperature of $T=\SI{450}{\MeV}$. We show results for the different angular momentum cutoffs $l_\text{max}=3,6,9$. Dot-dashed lines correspond to the respective overlaps in the Gibbs state.}
    \label{fig:1_QCDNLO450}
\end{figure}

In figure~\ref{fig:1_QCDNLO450}, we show the time evolution of the overlaps $O_n$ with the bottomonium eigenstates $nS=1S,2S,3S$. In all cases, we can see the emergence of a clear steady state around \SI{20}{\femto\meter}. We note that the dips in the overlaps before \SI{5}{\femto\meter} appear due to the rescattering of the wave function at the boundary of the box in the position space of our simulation. After the scattering at the boundary, the density at intermediate radii $r$ increases again, leading to an increase in the overlaps after the initial minimum. The $1S$ overlaps do not show such a minimum since the $1S$ eigenfunction has such a small radius that even after the rescattering, the density stays small at the radii where the $1S$ wave function is large. Therefore, the $1S$ overlaps do not increase after the scattering. As emphasized before, this rescattering will always happen in long-time simulations with a finite simulation volume; however, we stress that in phenomenological simulations, it does not have a strong impact due to larger simulation volumes, the short simulation times, and higher $l_\text{max}$ choices~\cite{Omar:2021kra,Brambilla:2022ynh}. We further observe that a higher $l_\text{max}$ generally leads to smaller overlaps, which is expected since a larger angular momentum space will, in the long run, lead to less occupation of the $l=0$ state, which directly leads to smaller $S$-wave overlaps. 

\begin{figure}[h]
    \centering
    \includegraphics[width=0.95\textwidth]{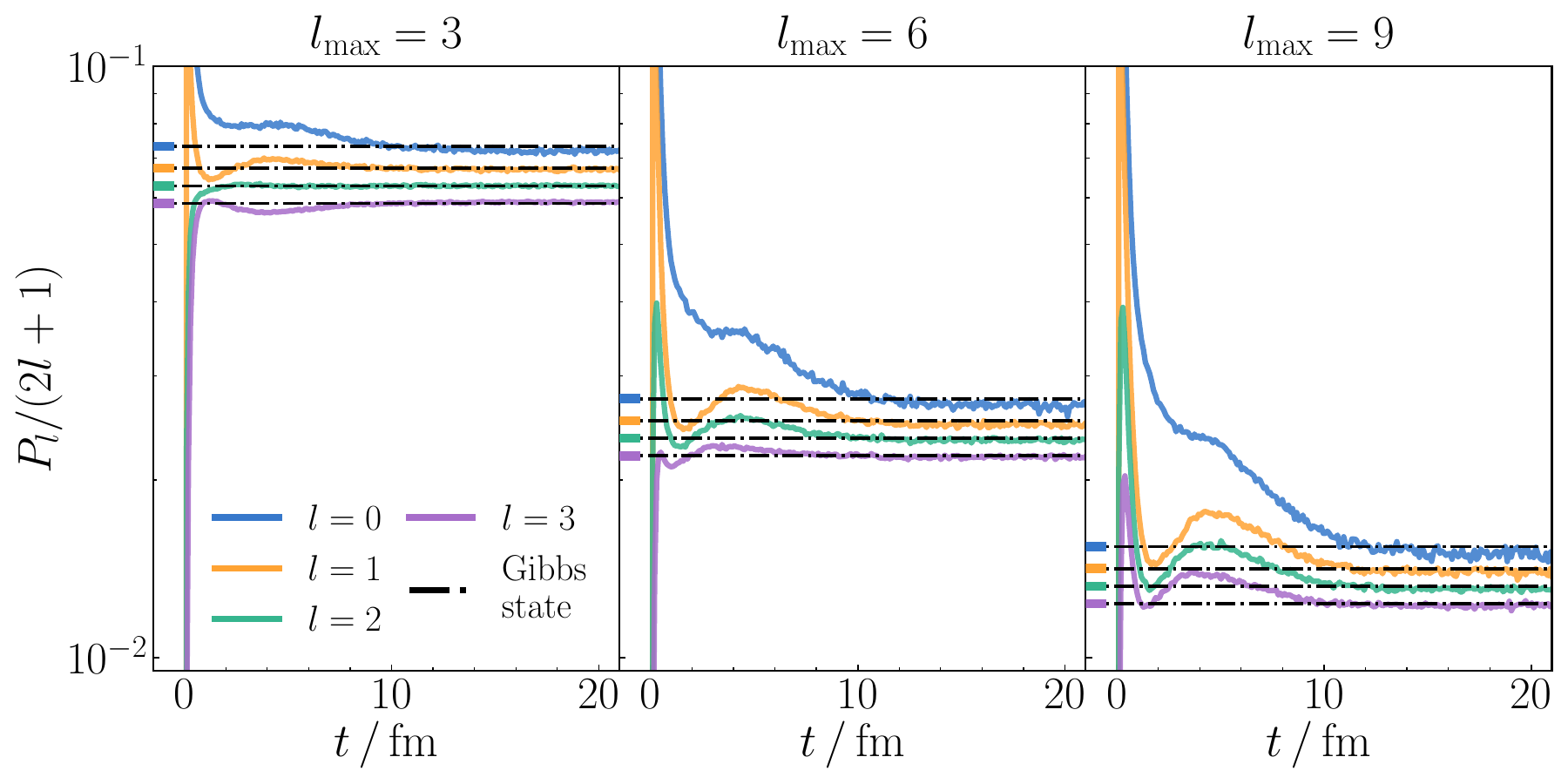}
    \caption{The angular momentum occupation divided by multiplicity, $P_l/(2l+1)$, from the simulation of the pNRQCD master equation at NLO in $E/(\pi T)$ at a constant temperature of $T=\SI{450}{\MeV}$. We show results for the different angular momentum cutoffs $l_\text{max}=3,6,9$. Dot-dashed lines correspond to the respective occupations in the Gibbs state.}
    \label{fig:2_QCDNLO450_l}
\end{figure}

In figure~\ref{fig:2_QCDNLO450_l}, we show the time evolution of the angular momentum occupation $P_l$ divided by the multiplicity $(2l+1)$. We observe equilibration in angular momentum space between \SI{10}{\femto\meter} and \SI{15}{\femto\meter}. A higher $l_\text{max}$ leads to a longer equilibration timescale since the angular momentum space to explore is larger; however, for the $l_\text{max}$ values considered here, the change in the equilibration timescale is marginal. 

\begin{figure}[h]
    \centering
    \includegraphics[width=0.9\textwidth]{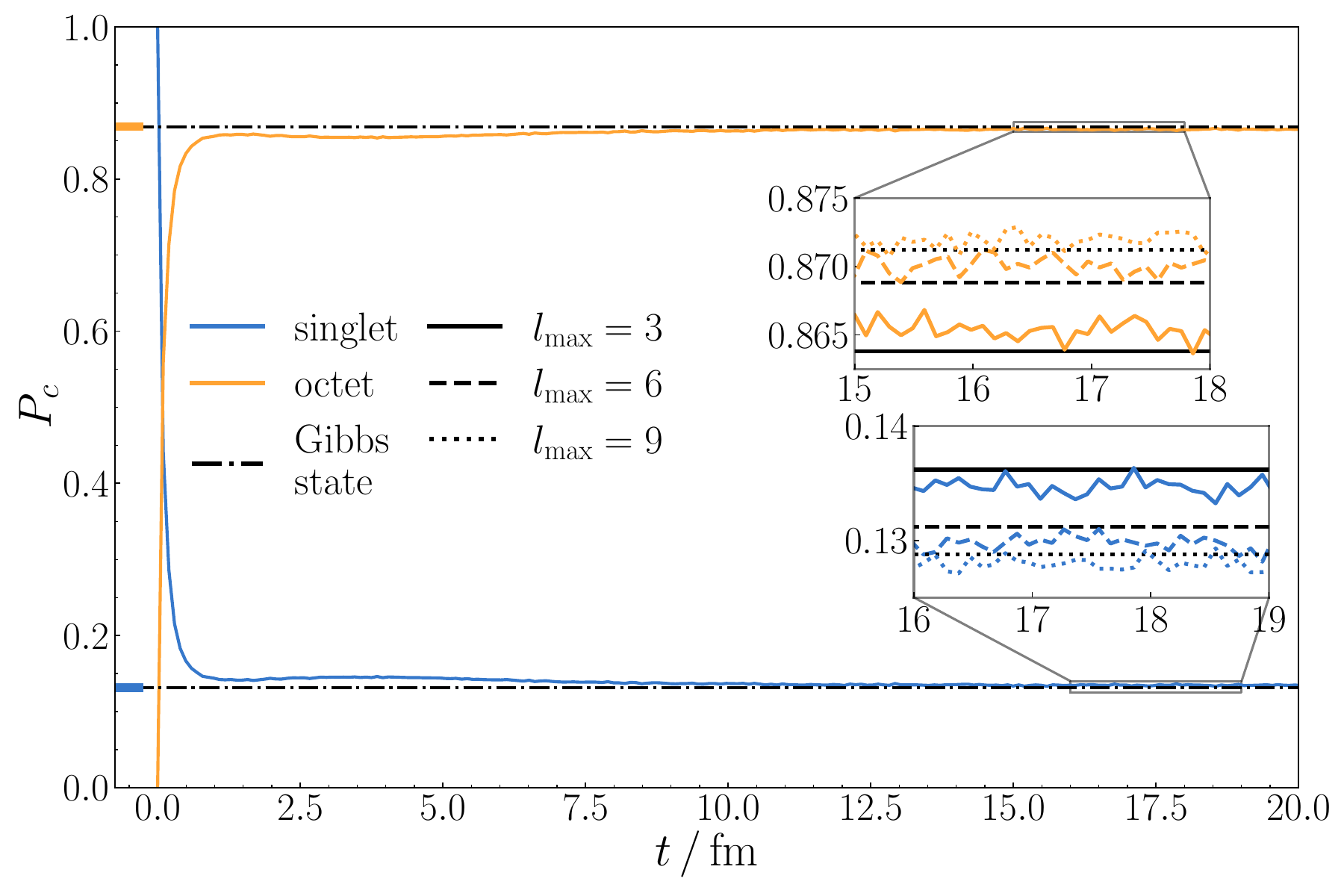}
    \caption{The color occupation $P_c$ from the simulation of the pNRQCD master equation at NLO in $E/(\pi T)$ at a constant temperature of $T=\SI{450}{\MeV}$. We show results for the different angular momentum cutoffs $l_\text{max}=3,6,9$. The black lines correspond to the respective occupations in the Gibbs state.}
    \label{fig:3_QCDNLO450_c}
\end{figure}

In figure~\ref{fig:3_QCDNLO450_c}, we display the evolution of the color occupation $P_c$. Equilibration in color space happens at about $\SI{10}{\femto\meter}-\SI{15}{\femto\meter}$ as well. Here, the angular momentum cutoff does not have a considerable effect on the timescale.

\begin{figure}[h]
    \centering
    \includegraphics[width=0.95\textwidth]{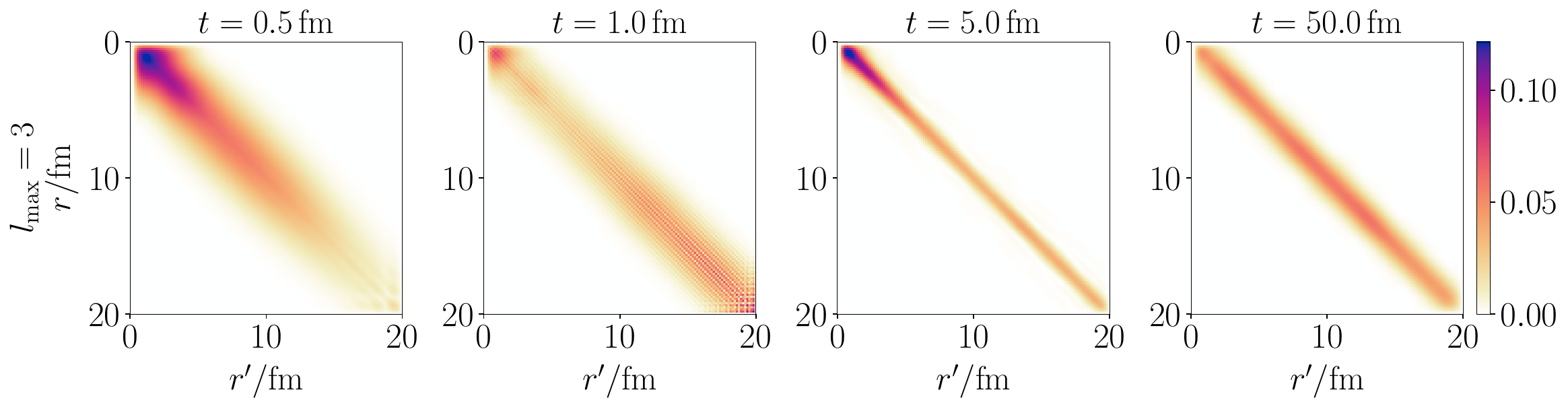}\\
    \includegraphics[width=0.95\textwidth]{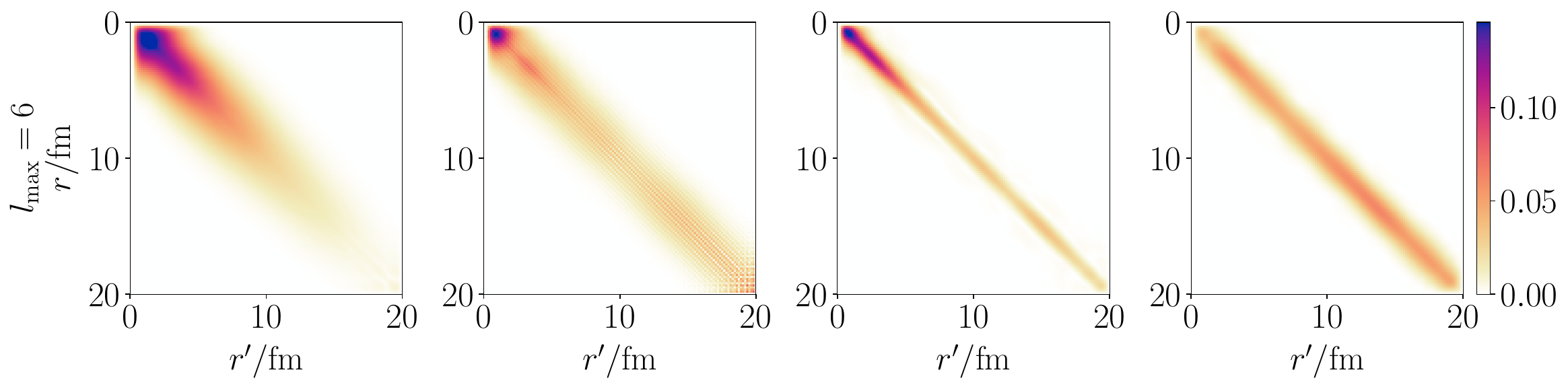}\\
    \includegraphics[width=0.95\textwidth]{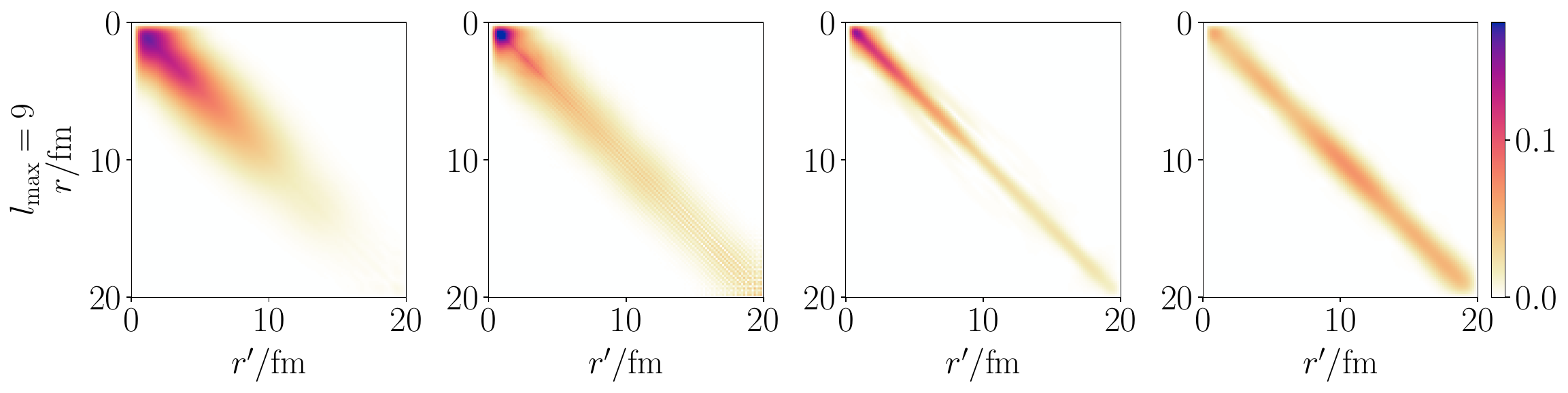}\\
    \caption{The absolute value of the singlet position space density matrix $|\rho_s(r,r^\prime)|$ projected on zero angular momentum $l=0$ for the simulation of the pNRQCD master equation at NLO in $E/(\pi T)$ at a constant temperature of $T=\SI{450}{\MeV}$. The columns correspond to different times $t$, ordered from \textit{left} to \textit{right}: \SI{0.5}{\femto\meter}, \SI{1}{\femto\meter}, \SI{5}{\femto\meter}, \SI{50}{\femto\meter}. The rows show results for different $l_\text{max}$, from \textit{top} to \textit{bottom}: $l_\text{max}=3$, $l_\text{max}=6$, $l_\text{max}=9$.}
    \label{fig:4_lmaxposition}
\end{figure}

In figure~\ref{fig:4_lmaxposition}, we show the evolution of the density matrix in position space, $\rho^{l=0}_s(r,r^\prime)$, where we project out singlet states with an angular momentum of $l=0$. We observe an initial density spread followed by the scattering at the boundary starting from $\SI{1}{\femto\meter}$ and a subsequent equilibration towards the steady state. We checked that the density matrix at later times is equal to the one observed at $\SI{50}{\femto\meter}$, confirming the equilibration in position space. In general, in position space, off-diagonal elements are suppressed, leading to a band with a certain width. The position space evolution at different cutoffs $l_\text{max}$ is very similar, however, we note that the rescattering at the bottom right boundary of the simulation domain at $t=\SI{1}{\femto\metre}$ is stronger for smaller $l_\text{max}$. This can be understood by considering the mixing with higher angular momentum states. For $l_\text{max}=3$, due to the small angular momentum space, states that are excited to higher angular momenta have a high probability to go back to $l=0$ very quickly. Since high angular momentum states have a more repulsive potential due to the centrifugal barrier $l(l+1)/(Mr^2)$, they contribute to pushing the density away from the origin. For larger $l_\text{max}$, since a state of, e.g., $l=3$ has a higher probability of being excited to even larger angular momenta, it takes a longer time until high angular momenta contribute to the $l=0$ position space distribution. 
This can also be seen in the overlaps in figure~\ref{fig:1_QCDNLO450}, where the dip due to rescattering for the $2S$ and $3S$ overlaps is later for larger $l_\text{max}$. 
Qualitatively, the evolution of the density matrix looks similar to previous one-dimensional simulations of quarkonium master equations~\cite{Miura:2022arv,Delorme:2024rdo}, which feature a similar spread of the density with a more localized peak towards $r=0$ due to the attractive singlet potential. However, the contributions of the repulsive angular momentum terms lead to enhancing the spread of the density and especially to the maximum in position space being located at larger $r$ at later times of the evolution.

\begin{figure}[h]
    \centering
    \includegraphics[width=0.95\textwidth]{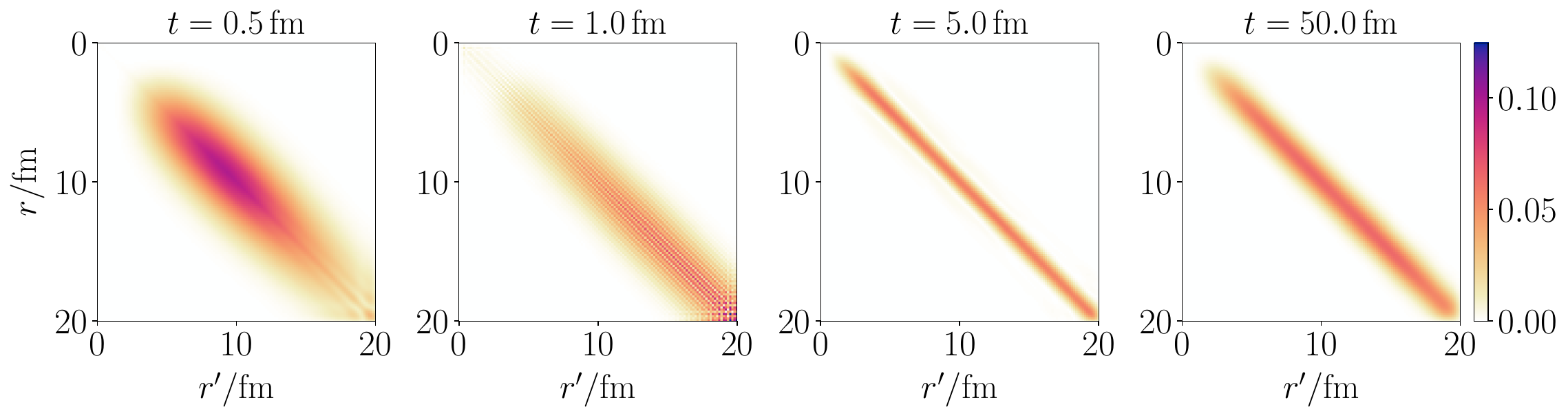}\\
    \includegraphics[width=0.95\textwidth]{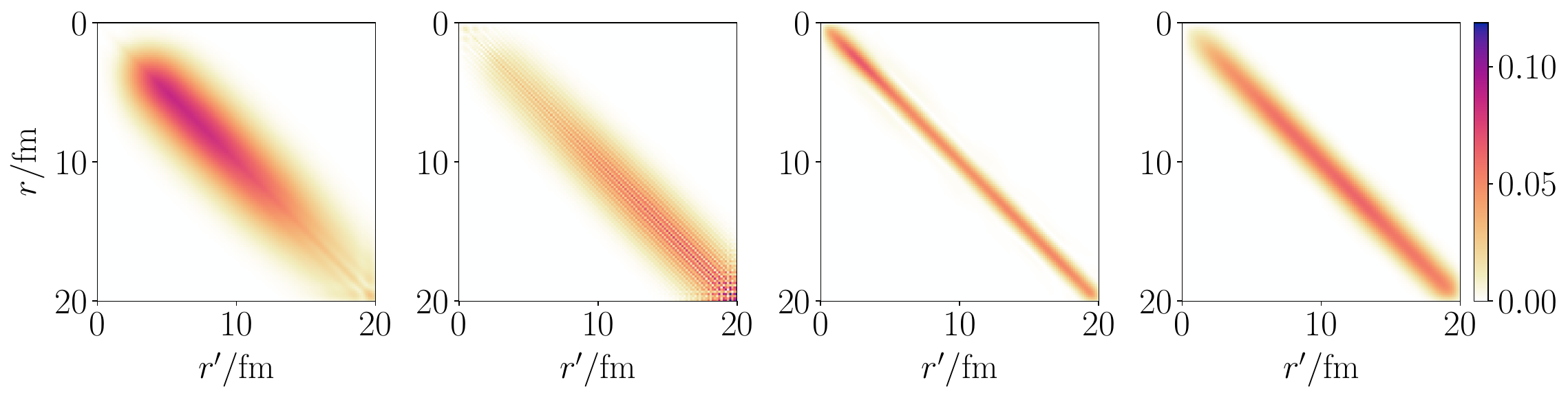}
    \caption{The absolute value of the position space density matrix $|\rho_c(r,r^\prime)|$ from the simulation of the pNRQCD master equation at NLO in $E/(\pi T)$ at a constant temperature of $T=\SI{450}{\MeV}$. \textit{Top:} The density matrix in the singlet sector $c=s$ projecting out the angular momentum $l=3$. \textit{Bottom:} The density matrix in the octet sector $c=o$ projecting out the angular momentum $l=0$. The columns correspond to different times $t$, ordered from \textit{left} to \textit{right}: $\SI{0.5}{\femto\meter}$, $\SI{1}{\femto\meter}$, $\SI{5}{\femto\meter}$, $\SI{50}{\femto\meter}$.}
    \label{fig:5_octetposition}
\end{figure}

In figure~\ref{fig:5_octetposition}, we furthermore show the time evolution of the position space density matrix when projecting out the singlet $l=3$ states and the octet $l=0$ states. We can see that in both these cases, the density is way less localized towards $r=0$, and stronger scattering takes place. In the octet case, this is due to the repulsive Coulomb potential $V_o(r)$ in the octet sector. In the singlet case with an angular momentum of $l=3$, this is due to the contribution of the centrifugal barrier, as discussed before.

\paragraph{Medium coupling dependence.}

We analyze the dependence on the coupling to the medium. As the transport coefficient $\gamma$ only leads to an effective modification of the potential and does not enter the Lindblad operators, we choose $\gamma=0$ and the influence of the medium is consequently only given by the transport coefficient $\kappa = \hat\kappa T^3$. Since $\hat\kappa$ is a property of the medium, a vanishing $\hat\kappa$ would resemble the evolution of the quarkonium in the vacuum. 
We simulate the NLO in $E/(\pi T)$ pNRQCD master equation for $\hat\kappa=4,2,0.5$ while keeping the temperature fixed at $T=\SI{450}{\MeV}$. To conserve computational resources, we restrain to simulations with $l_\text{max}=3$, where a time-step of $\delta t =\SI{0.00025}{\femto\meter}$ is permissible (see appendix~\ref{app:dt}). 

\begin{figure}[h]
    \centering
    \includegraphics[width=0.85\textwidth]{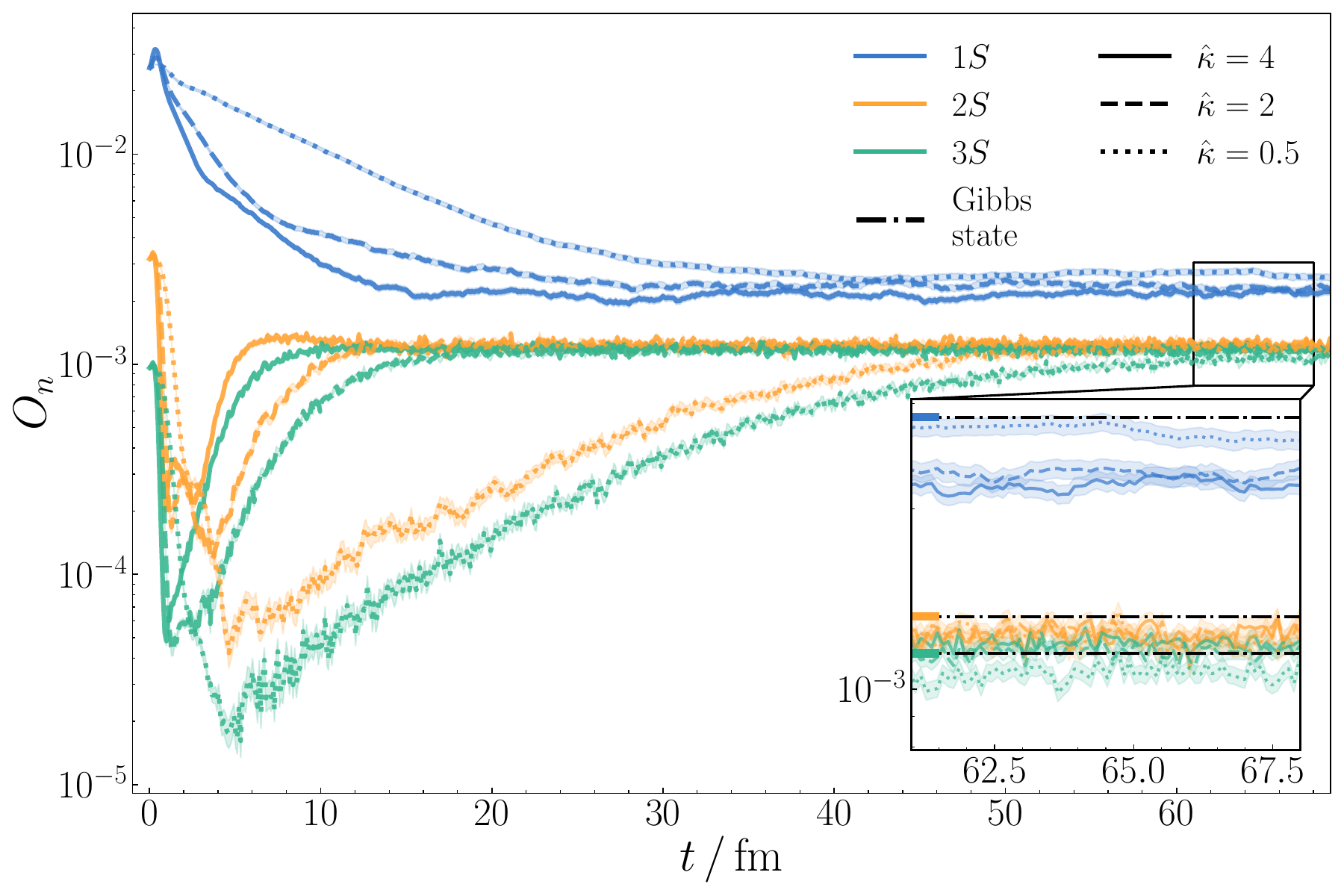}
    \caption{The time evolution of $S$-wave overlaps $O_n$ from a simulation of the pNRQCD master equation at NLO in $E/(\pi T)$ at fixed temperature $T=\SI{450}{\MeV}$. We compare transport coefficient values $\hat\kappa=4$ (solid), $\hat\kappa=2$ (dashed) and $\hat\kappa=0.5$ (dotted). The dot-dashed lines correspond to the respective overlaps in the Gibbs state.}
    \label{fig:6_QCDNLOkappa}
\end{figure}

The results for the evolution of the overlaps are shown in figure~\ref{fig:6_QCDNLOkappa}. We observe that a decreasing $\hat{\kappa}$ leads to a longer equilibration timescale, with the $1S$ overlap for $\hat{\kappa}=0.5$ only reaching a plateau around $\SI{40}{\femto\metre}-\SI{50}{\femto\meter}$. Notably for such small $\hat{\kappa}$, the $2S$ and $3S$ overlaps equilibrate even later, and it is unclear if the full steady state is already reached at $\SI{70}{\femto\meter}$. This increase in the timescale is expected, since, given $\kappa=\hat{\kappa}T^3$, if $\hat{\mathcal{L}} \hat{\rho} \propto\kappa$, the spectrum of $\hat{\mathcal{L}}$ scales linearly with $\kappa$. Therefore, the Liouvillian gap will also be linear in $\kappa$, leading to the thermalization timescale being proportional to $1/\kappa$. In our case, we still have the unitary term $-i[H,\rho]$, which does not depend on $\kappa$; however, it only contributes with an imaginary eigenvalue. Hence, at leading order in $\kappa$, the Liouvillian gap is still proportional to $\kappa$, leading to the observed behavior of the thermalization time scale. Notably, this differs from previous findings in the open Schwinger model, where, due to a strong and local dissipator, the thermalization timescale was found to increase with increasing dissipator strength~\cite{Angelides:2025hjt}. In kinetic theory, one finds a scaling law similar to ours for the kinetic equilibration time of the center-of-mass momentum $\tau_\text{kin}$, $\tau_\text{kin}\propto D_s$, where $D_s$ is the spatial diffusion coefficient~\cite{Svetitsky:1987gq,Moore:2004tg}. Since $D_s\propto 1/\kappa$ it follows that $\tau_\text{kin}\propto 1/\kappa$.

\begin{figure}[h]
    \centering
    \includegraphics[width=0.75\textwidth]{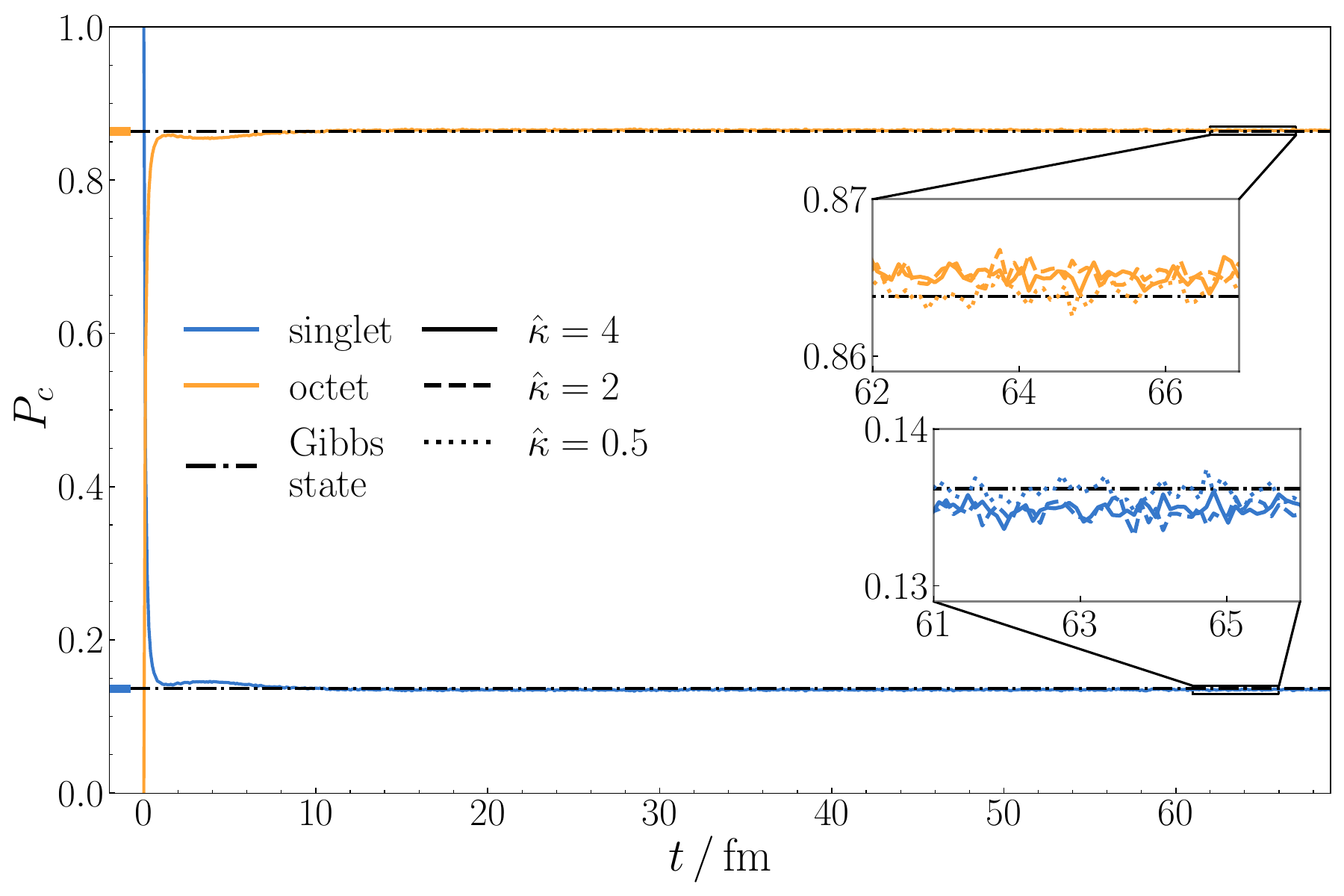}
    \caption{The time evolution of the color occupation $P_c$ from a simulation of the pNRQCD master equation at NLO in $E/(\pi T)$ at fixed temperature $T=\SI{450}{\MeV}$. We compare three transport coefficient values $\hat\kappa=4$ (solid), $\hat\kappa=2$ (dashed) and $\hat\kappa=0.5$ (dotted). The dot-dashed lines in the zoomed-in plots show the predictions for the Gibbs state.}
    \label{fig:8_QCDNLOkappa_c}
\end{figure}

In figure~\ref{fig:8_QCDNLOkappa_c}, we show the evolution of the color occupation for three different transport coefficient values.
The equilibration time in color space is only mildly affected by $\hat\kappa$, with the steady state still emerging at around $\SI{10}{\femto\metre}$.

\paragraph{Temperature dependence}

To examine the influence of the plasma's temperature on the equilibration process, we compare the evolution of bottomonium in a medium with temperature $T=\SI{450}{\MeV}$ against a medium with temperature $T=\SI{250}{\MeV}$. We again restrict ourselves to $l_\text{max}=3$ and consider the pNRQCD master equation at NLO in $E/(\pi T)$. We note that, since $\kappa=\hat\kappa T^3$, in the case of a lower temperature, the coupling to the environment is also smaller. 

\begin{figure}[h]
    \centering
\includegraphics[width=0.75\textwidth]{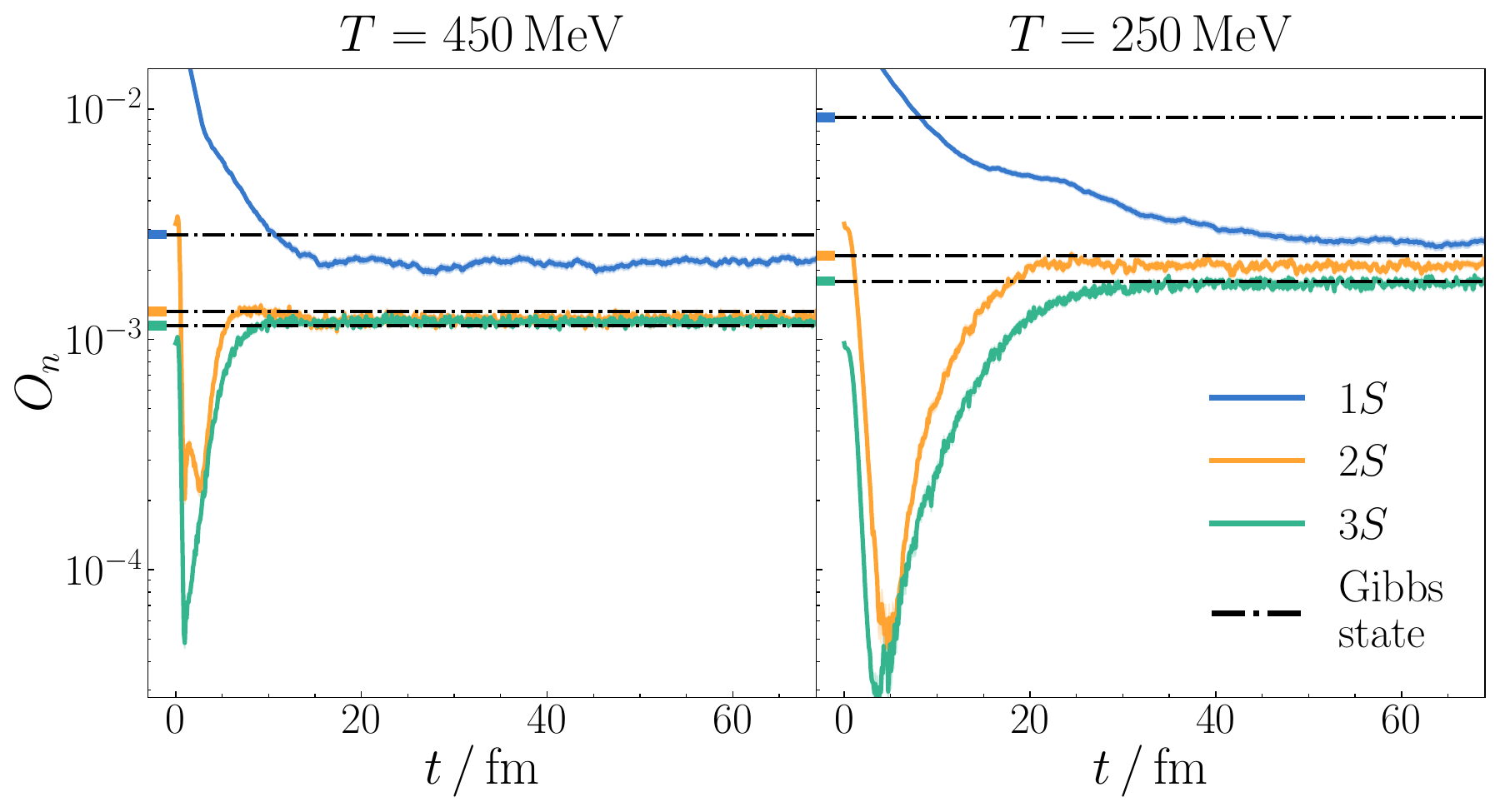}
    \caption{The time evolution of $S$-wave overlaps $O_n$ from a simulation of the pNRQCD master equation at NLO in $E/(\pi T)$. We show results for the two different temperatures $T=\SI{450}{\MeV}$ \textit{(left)} and $T=\SI{250}{\MeV}$ \textit{(right)}. The dot-dashed lines correspond to the respective overlaps in the Gibbs state.}
    \label{fig:9_QCDNLO250}
\end{figure}

In figure~\ref{fig:9_QCDNLO250}, we compare the evolution of the overlaps for the two different temperatures.
We observe that for the lower temperature $T=\SI{250}{\MeV}$ the equilibration takes longer and the $1S$ overlap only plateaus at $\SI{50}{\femto\metre}-\SI{60}{\femto\meter}$. In kinetic theory, the equilibration time of the center-of-mass momentum obeys a similar scaling law~\cite{Scardina:2017ipo}.

\begin{figure}[h]
    \centering
\includegraphics[width=0.75\textwidth]{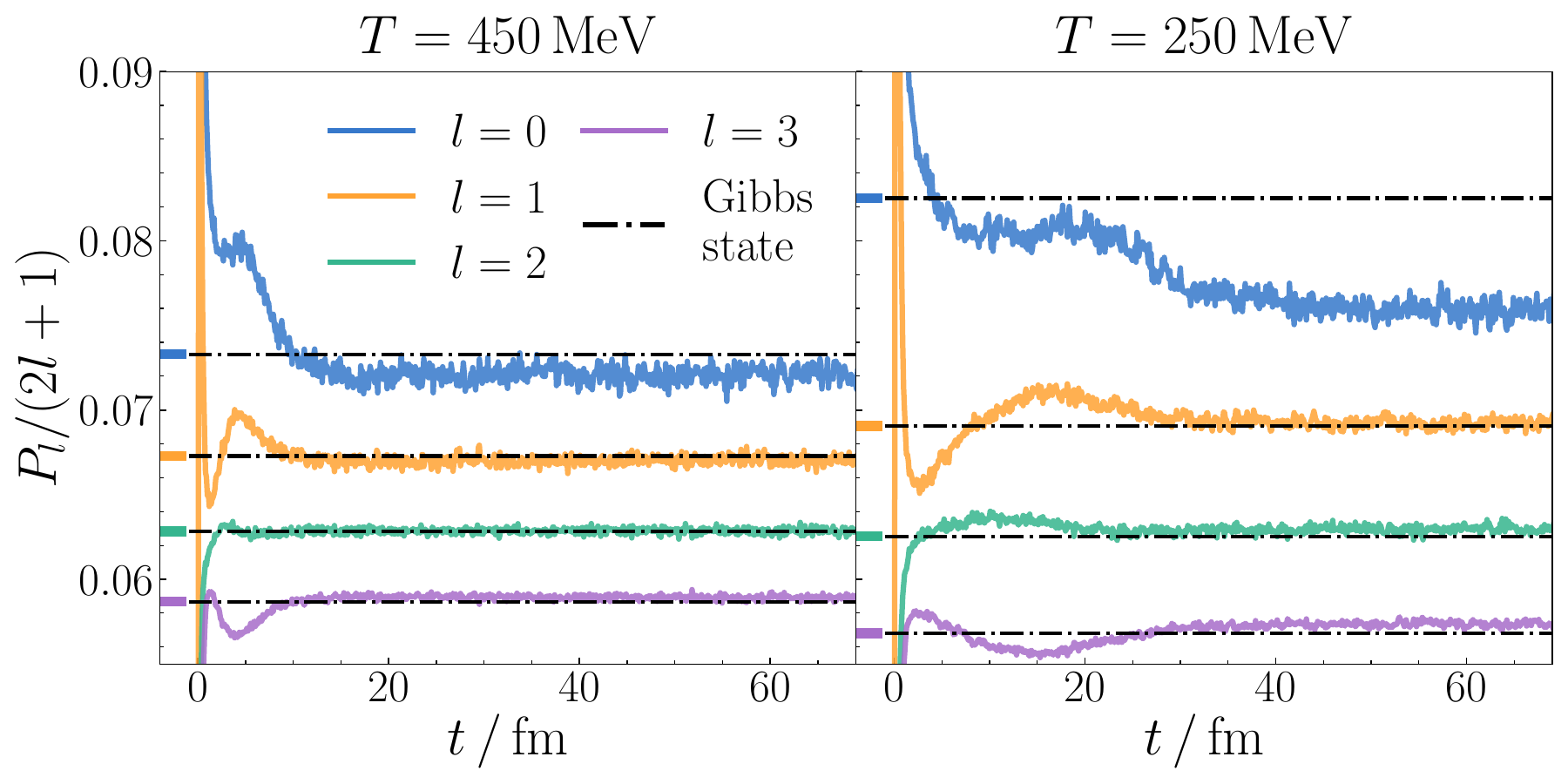}    
    \caption{The angular momentum occupation divided by multiplicity $P_l/(2l+1)$ from a simulation of the pNRQCD master equation at NLO in $E/(\pi T)$. We compare the two temperatures $T=\SI{450}{\MeV}$ \textit{(left)} and $T=\SI{250}{\MeV}$ \textit{(right)}. Dot-dashed lines show the predictions for the Gibbs state.}
    \label{fig:10_QCDNLO250_l}
\end{figure}

\begin{figure}[h]
    \centering
\includegraphics[width=0.75\textwidth]{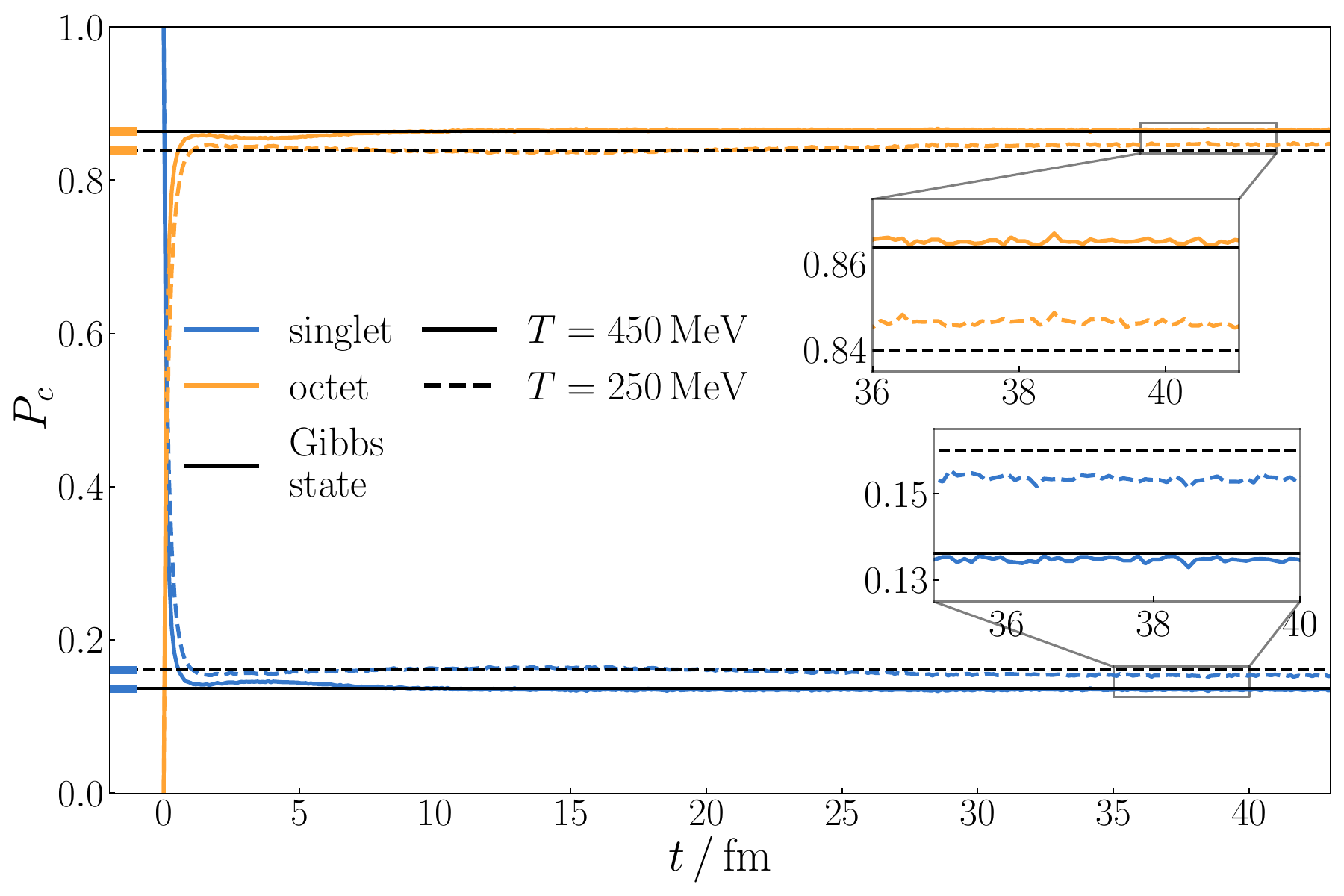}    
    \caption{The color occupation $P_c$, from a simulation of the pNRQCD master equation at NLO in $E/(\pi T)$. We compare the two temperatures $T=\SI{450}{\MeV}$ (solid) and $T=\SI{250}{\MeV}$ (dashed). The horizontal black lines show the predictions for the Gibbs state.}
    \label{fig:11_QCDNLO250_c}
\end{figure}

In figure~\ref{fig:10_QCDNLO250_l}, we compare the evolution in angular momentum for both temperatures. We find that for lower temperatures, equilibration takes longer, with the steady state in angular momentum space being reached at about $\SI{40}{\femto\metre}$ in the case of $T=\SI{250}{\MeV}$. 
In figure~\ref{fig:11_QCDNLO250_c}, we show the evolution of the color occupation.
In color space, thermalization also takes longer for lower temperatures, with equilibrium emerging at approximately $\SI{30}{\femto\metre}$.

\begin{figure}[h]
    \centering
    \includegraphics[width=0.95\textwidth]{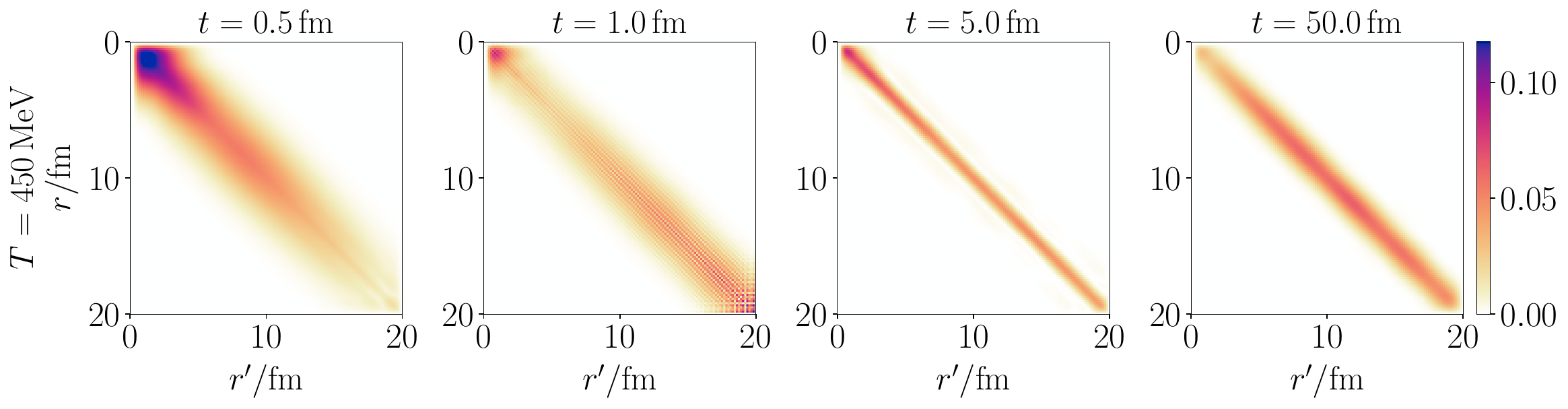}\\
    \includegraphics[width=0.95\textwidth]{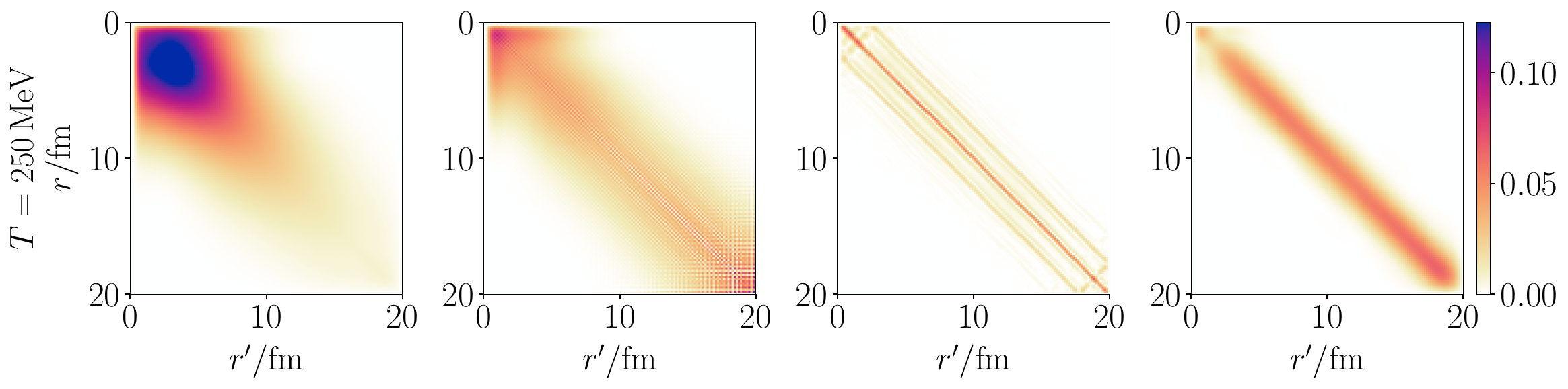}
    \caption{The absolute value of the color singlet position space density matrix $|\rho_s(r,r^\prime)|$ projected on the angular momentum $l=0$ from a simulation of the pNRQCD master equation at NLO in $E/(\pi T)$. \textit{Top:} The evolution for a medium temperature of $T=\SI{450}{\MeV}$ \textit{Bottom:} The evolution for a medium temperature of $T=\SI{250}{\MeV}$. The columns correspond to different times $t$, ordered from \textit{left} to \textit{right}: \SI{0.5}{\femto\meter}, \SI{1}{\femto\meter}, \SI{5}{\femto\meter}, \SI{50}{\femto\meter}.}
    \label{fig:12_T250positionspace}
\end{figure}

Finally, in figure~\ref{fig:12_T250positionspace}, we compare the evolution in position space between the two temperatures. For the bottom row showing the $T=\SI{250}{\MeV}$ evolution, we see that the density has a larger width around the diagonal. Off-diagonal elements in position space are therefore suppressed by the temperature. At $\SI{5}{\femto\metre}$, the $T=\SI{250}{\MeV}$ density matrix seems to be more diagonal than the $T=\SI{450}{\MeV}$ one; however, this is due to the longer thermalization timescale for the lower temperature. In general, the density matrix takes a very diagonal form right after the rescattering process. In the case of $T=\SI{250}{\MeV}$, it takes a longer time until the rescattering is completed, which here only happens at $\SI{5}{\femto\metre}$. At this point, the $T=\SI{450}{\MeV}$ density matrix is less diagonal only because its rescattering process happened before. 

\paragraph{pNRQCD at LO in $E/(\pi T)$.}

\begin{figure}[h]
    \centering
    \includegraphics[width=0.95\textwidth]{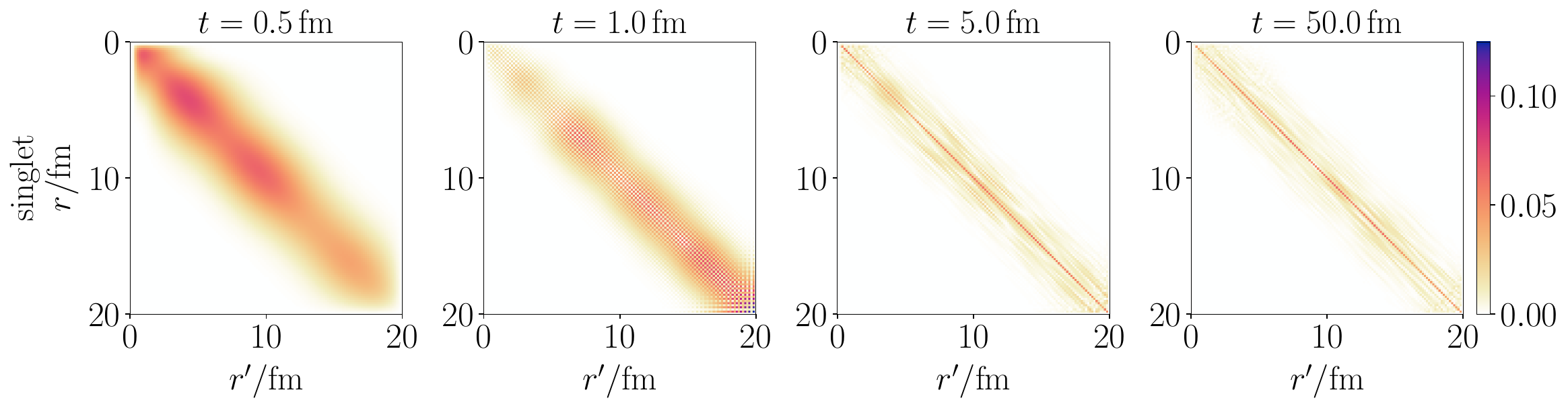}
    \includegraphics[width=0.95\textwidth]{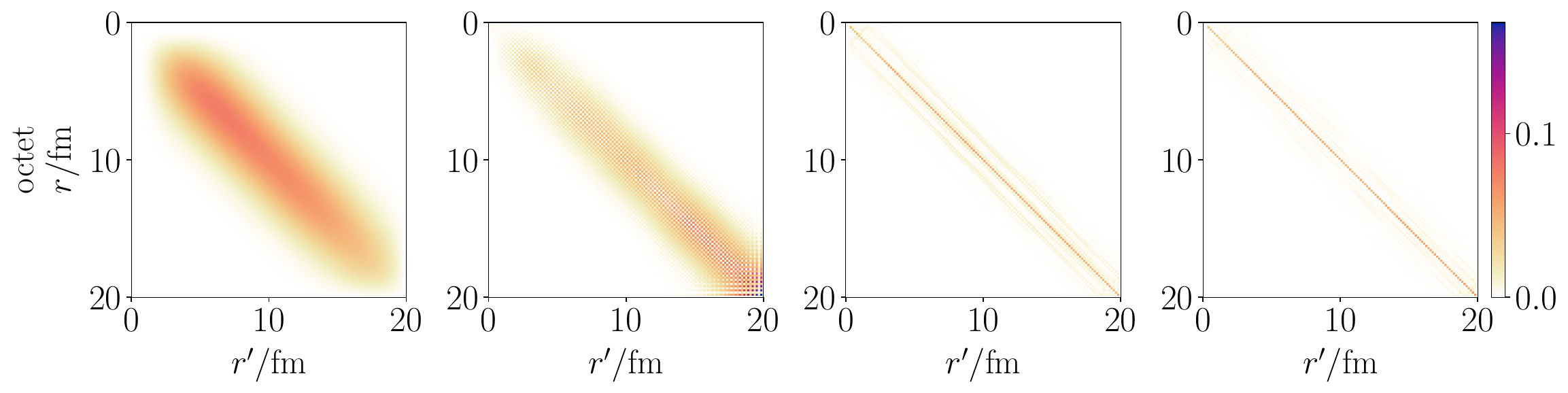}
    \caption{The absolute value of the position space density matrix $|\rho_c(r,r^\prime)|$ from a simulation of the pNRQCD master equation at LO in $E/(\pi T)$. \textit{Top:} The density matrix in the singlet sector $c=s$ projecting out the angular momentum $l=0$. \textit{Bottom:} The density matrix in the octet sector $c=o$ projecting out the angular momentum $l=0$. The columns correspond to different times $t$, ordered from \textit{left} to \textit{right}: \SI{0.5}{\femto\meter}, \SI{1}{\femto\meter}, \SI{20}{\femto\meter}, \SI{50}{\femto\meter}.}
    \label{fig:13_density_LO}
\end{figure}

To estimate the impact of the next-to-leading order terms in the $E/(\pi T)$ expansions, we perform a simulation of the pNRQCD master equation at leading order in $E/(\pi T)$ following from eqs.~\eqref{eq:HamiltonianLO}-\eqref{eq:C1QCDLO}.
The evolution of the position space density matrix for $l=0$ in both the singlet and octet sector is shown in figure~\ref{fig:13_density_LO}.
We observe that at $\SI{1}{\femto\metre}$, there is a strong scattering at the boundary, without the characteristic maximum at small $r$, which we observed in the NLO case. Afterward, the density approaches a diagonal form. This is quite different from what we observed in the previous sections. However, we can understand it in this case by looking at the Lindblad equation analytically. We find that we can write down a closed form for the exact steady state as the density matrix
\begin{equation}
    \rho_\text{LO} = \begin{pmatrix}
        \rho_s & 0\\
        0 & \rho_o 
    \end{pmatrix} = \begin{pmatrix}
        \mathds{1} & 0 \\
        0 & (N^2_c - 1)\mathds{1}
    \end{pmatrix}.
    \label{eq:SSLO}
\end{equation}
We can verify that this is the exact steady state by explicitly calculating $d\rho(t)/dt=\mathcal{L}[\rho(t)]$, which turns out to vanish for $\rho_\text{LO}$. Assuming the uniqueness of the steady state for the LO pNRQCD master equation, for which we have provided arguments, the system has to evolve towards this state. In angular momentum space, the system quickly relaxes to the expected values for $\rho_\text{LO}$, $P_l=2l+1$, within about $\SI{2}{\femto\metre}$ and in color space to $P_s=1/9$ and $P_o=8/9$ within $\SI{5}{\femto\metre}$. In the LO case, the system, therefore, equilibrates faster than in the NLO case. We also note that the steady state color occupation corresponds exactly to the prediction made in the static limit~\cite{Brambilla:2017zei}.

\begin{figure}[h]
    \centering
        \includegraphics[width=0.85\textwidth]{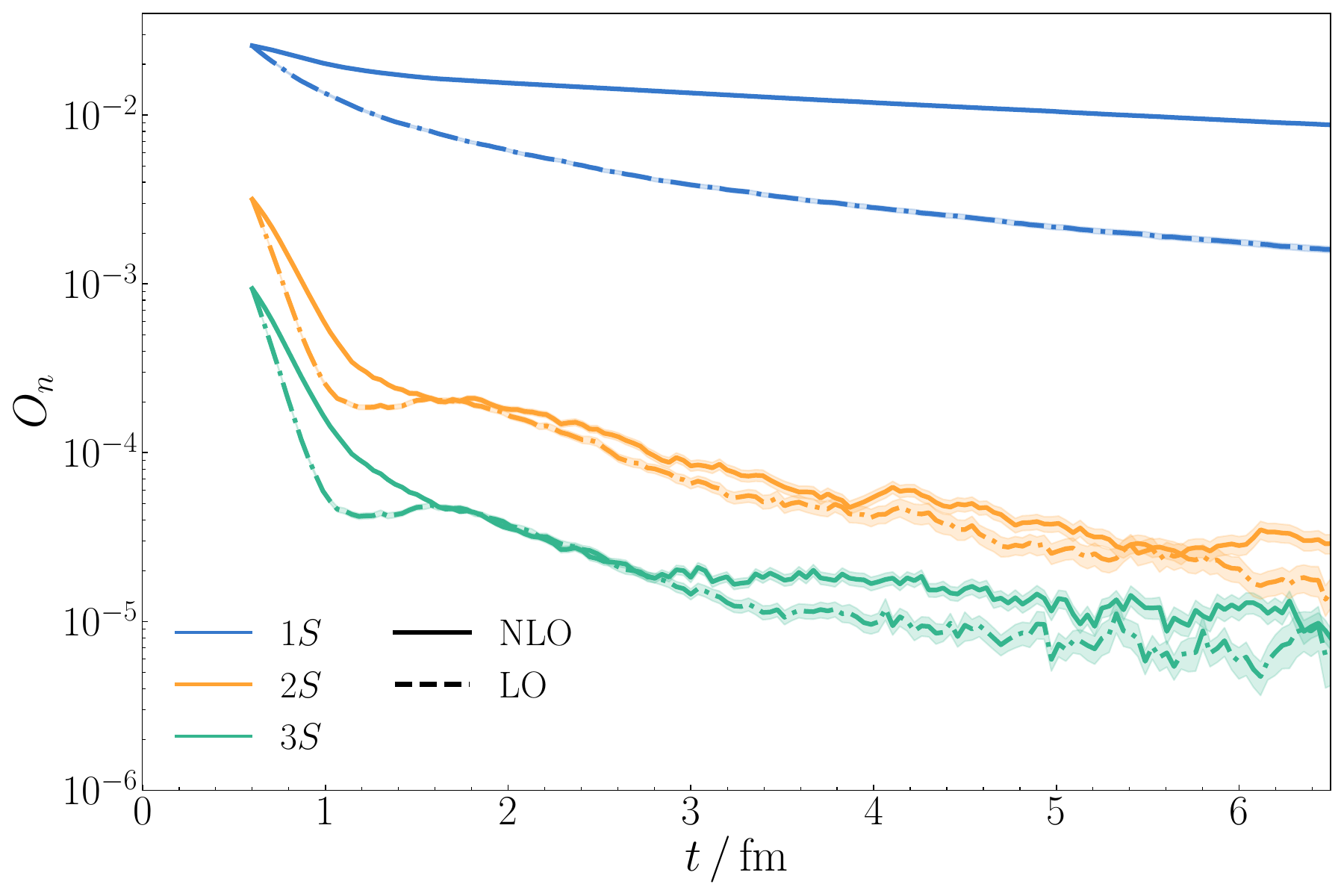}    
    \caption{The overlaps $O_n$ for the $1S,2S$ and $3S$ quarkonium states. We compare simulations of the pNRQCD master equation at LO and NLO in $E/(\pi T)$ for a medium with a Bjorken temperature evolution from $T=\SI{425}{\MeV}$ to $\SI{190}{\MeV}$.}
    \label{fig:16_LOpheno}
\end{figure}

While for the long-time behavior, the NLO terms in the master equation seem to play a critical role, for the evolution at early times up to $\SI{10}{\femto\metre}$, relevant for phenomenology, and especially at lower temperatures, the next to leading order terms give sizable, but reasonable corrections to the evolution. To show this, in figure~\ref{fig:16_LOpheno} we compare the time evolution of the pNRQCD master equation at LO and NLO in $E/(\pi T)$. In both cases, we perform simulations with parameters close to the ones used in phenomenological studies of quarkonium suppression, by employing a Bjorken temperature evolution~\cite{Bjorken:1982qr} from $\SI{425}{\MeV}$ to $\SI{190}{\MeV}$ and not setting an $l_\text{max}$ cutoff. We observe that the evolution of the $2S$ and $3S$ states for the LO equation agrees reasonably well with the NLO results. The $1S$ overlaps, on the other hand, show stronger suppression compared to the NLO solution. We conclude that, while both equations qualitatively agree in the phenomenological case, the NLO terms add sizable corrections to the $1S$ overlaps.

\subsection{How does the steady state look like ?}
\label{sec:steady_state}

Finally, we characterize the structure of the steady state. As discussed in section~\ref{sec:sstates}, in general, we do not expect an open system to relax to the Gibbs state $\rho_\text{Gibbs}$. In fact, $\rho_\text{Gibbs}$ is not a solution for any of the master equations we are considering. However, in the case $\gamma=0$, we have 
\begin{equation}
    \mathcal{L}[\rho_\text{Gibbs}]\propto \kappa,
\end{equation}
and therefore, we expect the steady state to approach the Gibbs state for $\kappa\to 0$. For a finite $\kappa$, however, we expect corrections in $\kappa$. We estimate these corrections by comparing the steady state results with the expectations from the Gibbs state calculated as described in section~\ref{sec:timescale}.

\paragraph{pNRQCD at NLO.}

To assess the corrections to the Gibbs state for the NLO pNRQCD master equation, in figure~\ref{fig:1_QCDNLO450}, we compare the overlaps in the steady state to the predictions from the Gibbs state, indicated by the dot-dashed lines. We observe that the overlaps in the steady state, especially for the $1S$, consistently lie below the predictions from the Gibbs state. This is in agreement with previous findings for a one-dimensional master equation in ref.~\cite{Delorme:2024rdo}, where the authors also found that the overlaps of the bound states are lower than predicted by the Gibbs state. In addition, we observe that the ratio of the steady state $1S$ overlap and the Gibbs prediction stays approximately the same for increasing $l_\text{max}$. The fact that the difference between the steady state and the Gibbs state is larger for the $1S$ state than for the excited states may possibly originate from the larger $E/(\pi T)$ corrections in the case of the ground state.

\begin{figure}
    \centering
    \includegraphics[width=0.5\linewidth]{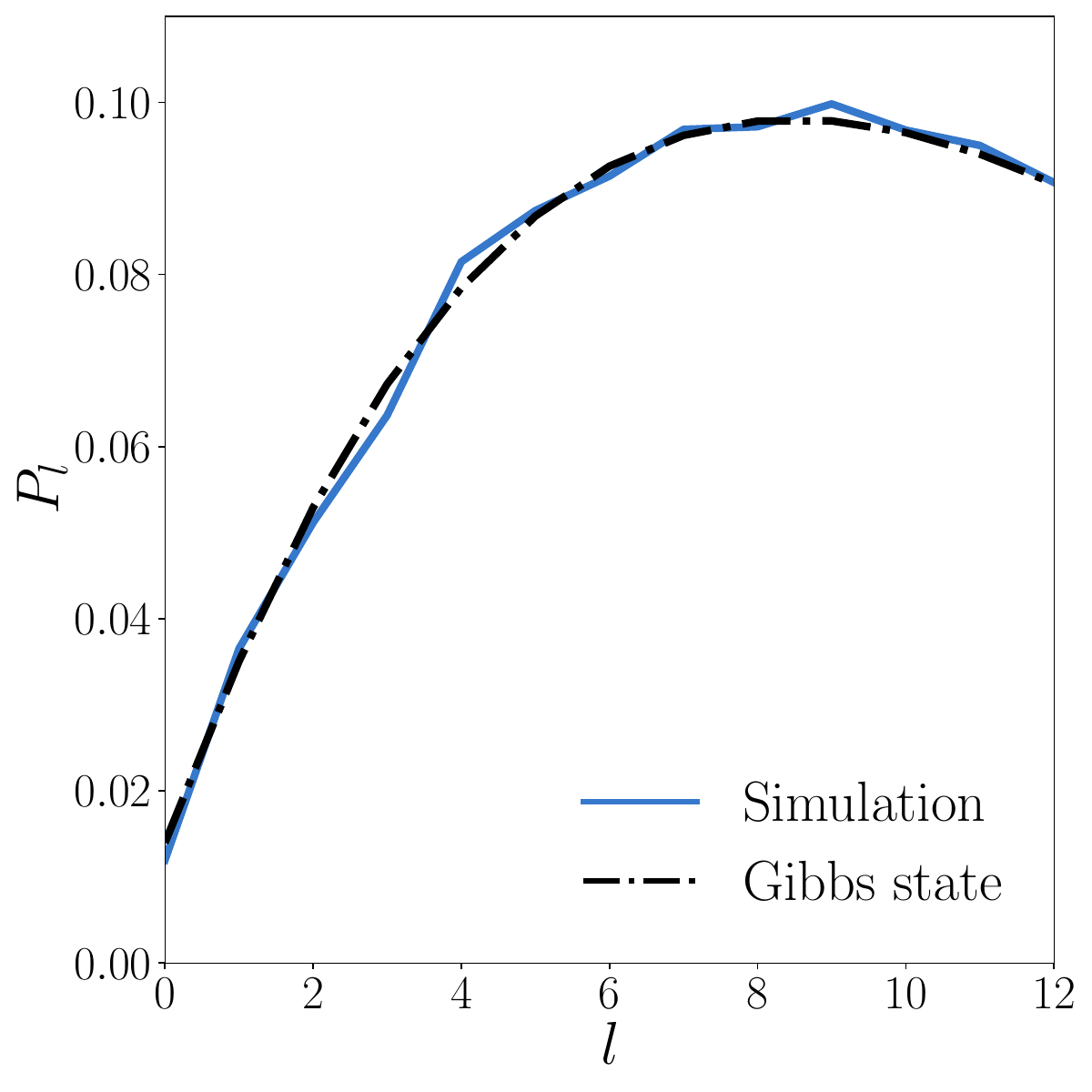}
    \caption{The angular momentum occupation $P_l$ at $t=\SI{70}{\femto\meter}$ from simulations of the NLO in $E/(\pi T)$ master equation at $T=\SI{250}{\MeV}$. The simulations are performed with a maximum angular momentum of $l_\text{max}=12$. The dot-dashed line shows the corresponding predictions for the Gibbs state.}
    \label{fig:0_lm12_Pl}
\end{figure}

Concerning the angular momentum occupation $P_l$, we observe in figure~\ref{fig:2_QCDNLO450_l} that for the first three angular momentum levels, and all $l_\text{max}$, the steady state shows very small corrections to the Gibbs state. As previously noted in section~\ref{sec:timescale}, we expect the Gibbs state to have a maximum in the angular momentum occupation $P_l$, and the next-to-leading order jump probabilities~\eqref{eq:nlopup} and~\eqref{eq:nlopdown} to be able to reproduce such a maximum. To test this, we perform a simulation of the next-to-leading order pNRQCD master equation with $l_\text{max}=12$ and a temperature of $T=\SI{250}{\MeV}$. Since smaller temperatures shift the maximum of $P_l$ in the Gibbs state to lower $l$, we should be able to observe a maximum in such simulations. We simulate until $\SI{70}{\femto\meter}$, where the steady is reached. The results for $P_l$ in the steady state are shown in figure~\ref{fig:0_lm12_Pl}. We observe that the steady state results agree well with the Gibbs state for all angular momenta. Especially, we note that the $P_l$ of the simulation shows a maximum at $l=9$, demonstrating the property of the NLO master equation to evolve towards a steady state with a $P_l$ that is not monotonically increasing.
For the color occupation $P_c$ in figure~\ref{fig:3_QCDNLO450_c}, we observe a similar agreement with the Gibbs state for the different $l_\text{max}$.

\paragraph{Transport coefficient dependence.}

Since for $\gamma=0$ all terms in $\mathcal{L}[\rho_\text{Gibbs}]$ are proportional to $\kappa$ we expect to approach the Gibbs state for $\kappa\to 0$.
We check this by comparing the steady state results for $\hat\kappa = 4,2,0.5$ to the respective Gibbs state predictions. In figure~\ref{fig:6_QCDNLOkappa}, we see that the $1S$ overlap approaches the value at the Gibbs state for decreasing $\hat\kappa$. The $3S$ overlap appears to be slightly below the Gibbs state prediction for $\hat{\kappa}=0.5$, however due to the substantially larger thermalization timescale for such a small $\hat{\kappa}$, it might be that the equilibrium is not fully reached yet. Similarly, in figure~\ref{fig:8_QCDNLOkappa_c}, the color occupations come slightly closer to the Gibbs state predictions for decreasing $\hat\kappa$.
These results confirm the physical intuition that with medium effects becoming small, the free evolution leads to thermalization with the bath.

\paragraph{Temperature dependence.}

We consider the temperature dependence of the steady state by comparing the results for $T=\SI{450}{\MeV}$ and $T=\SI{250}{\MeV}$. In figure~\ref{fig:9_QCDNLO250}, figure~\ref{fig:10_QCDNLO250_l}, and figure~\ref{fig:11_QCDNLO250_c}, we notice that for all observables, the steady state for $T=\SI{250}{\MeV}$ shows larger corrections than for $T=\SI{450}{\MeV}$. This is expected, since in the high temperature limit, due to the degeneracy of the octet sector, the Gibbs state approaches the density matrix given in eq.~\eqref{eq:SSLO}. At the same time, for high temperatures, the next-to-leading order terms become negligible, resulting in the leading order steady state~\eqref{eq:SSLO}. Therefore, deviation from the Gibbs state is always suppressed by the temperature.
Furthermore, in figure~\ref{fig:12_T250positionspace}, we see that the steady state for $T=\SI{250}{\MeV}$ has a wider width around the diagonal due to the weaker suppression of the off-diagonal elements.

\section{Conclusions}
\label{sec:conc}

In this work, we address the question of the thermalization of bottomonium inside a thermal medium in the open quantum system framework. We set out to determine (1) whether an equilibrium state exists, (2) the timescale of equilibration, and (3) the nature of the steady state.
Our findings support the following conclusions.
\begin{enumerate}
    \item The Lindblad equations considered in this study may have a unique and attractive steady state.
    \item The timescale of thermalization depends on the cutoff imposed on the Hilbert space. The timescale increases for decreasing temperature and decreasing medium coupling. Equilibration in color space takes place in approximately $\SI{10}{\femto\meter}$ and depends only weakly on the Hilbert space cutoff or medium parameters.
    \item The steady state shows corrections to the Gibbs state, which, however, vanish in the limit of vanishing coupling to the medium and large temperature, as expected. At a temperature of $T=\SI{450}{\MeV}$, the angular momentum and color occupation approximately thermalize to the Gibbs state, and the largest corrections are in the $1S$ overlap.
\end{enumerate}
In contrast to the findings of the second point, ref.~\cite{Angelides:2025hjt} showed that stronger dissipation actually slows down the equilibration in their open Schwinger model computation. This difference is likely due to the fact that they employ a very strong, local dissipator, and demonstrates that different models for the quarkonium evolution can show even qualitative differences in their equilibration behavior. 

Our results provide insight into the approach to equilibrium of quarkonium within the open quantum system approach in three dimensions, possibly contributing to a deeper understanding of quarkonium dynamics in this framework. In future work, the study of the Wigner distribution and entanglement of the system might further reveal how quantum features of the system are suppressed, as it approaches equilibrium, and at what times the system can effectively be described classically. Eventually, these insights could clarify the need for genuine quantum simulations of quarkonium suppression. Furthermore, it would be interesting to explore how the approach to equilibrium and steady-state properties change when the $E/(\pi T)$ expansion is not applied, which is one of the main approximations necessary to obtain a Lindblad equation. To extend this study to charmonium, where recombination effects are important, it would be necessary to develop an open quantum system formulation that correctly accounts for the regeneration of charmonium from initially uncorrelated charm quark pairs, which is one of the challenges ahead.

\acknowledgments

T.M. thanks Herbert Spohn for useful discussion. N.B., T.M., and A.V. further thank Jean-Paul Blaizot, Miguel Angel Escobedo, Arthur Lin, Michael Strickland, and Peter Vander Griend for their valuable comments. N.B., T.M., and A.V. acknowledge support by the DFG cluster of excellence ORIGINS funded by the Deutsche Forschungsgemeinschaft (DFG) under Germany's Excellence Strategy - EXC-2094-390783311. The authors gratefully acknowledge the Gauss Centre for Supercomputing e.V. (\href{www.gauss-centre.eu}{www.gauss-centre.eu}) for funding this project by providing computing time on the GCS Supercomputer SuperMUC-NG at the Leibniz Supercomputing Centre (\href{www.lrz.de}{www.lrz.de}).
N.B. acknowledges the European Research Council advanced grant  ERC-2023-ADG-Project EFT-XYZ. The work of N.B. is supported by the DFG Grant No. BR 4058/5-1 "Open quantum systems and effective field theories for hard probes of hot and/or dense medium".

\appendix
\section{Uniqueness of the steady state}
\label{app:proofs}

In this appendix, we explicitly show that there is no analytic operator in $p$ and $r$ satisfying \eqref{eq:cond1}-\eqref{eq:cond3} in one dimension that is not proportional to the identity.

\paragraph{pNRQCD at NLO in $E/(\pi T)$.}

Due to the color structure, we can write the operator $O$ as 
\begin{equation}
    O = \sum_{kl}\begin{pmatrix}c_{0kl} & c_{1kl}\\ c_{2kl} & c_{3kl}\end{pmatrix}r^kp^l,
    \label{eq:genopqcd}
\end{equation}
with coefficients $c_{ikl}$. We then have the conditions
\begin{gather}
\begin{split}
    [H,O]&=0,\\
    [C^{0(\dagger)},O]&=0,\\
    [C^{1(\dagger)},O]&=0,
\end{split}
\end{gather}
which need to be satisfied. For the steady state to be unique, we need to show that $O$ is proportional to $\mathds{1}$, i.e. that all $c_{ikl}$ vanish, except $c_{000}$ and $c_{300}$, which have to satisfy $c_{000}=c_{300}$.

To constrain the coefficients for the pNRQCD equation at NLO, we start from the conditions for $C^{1(\dagger)}$. We obtain three equations
\begin{align}
    \sum_{kl}\left(r^{k+1}p^l - ilr^kp^{l-1}\pm\frac{i}{2MT}r^kp^{l+1}\right)c_{1kl}=0,\label{eq:c11}\\
    \sum_{kl}\left(r^{k+1}p^l \pm \frac{i}{2MT}\left(r^kp^{l+1} - ikr^{k-1}p^l\right)\right)c_{2kl}=0,\label{eq:c12}\\
    \sum_{kl}\left[ilr^kp^{l-1}\pm \frac{1}{2MT}kr^{k-1}p^l\right]c_{3kl}=0.\label{eq:c13}
\end{align}
Adding and subtracting the different cases for $C^1$ and $C^{1\dagger}$ of eqs.~\eqref{eq:c11} and~\eqref{eq:c12} respectively gives the equations
\begin{align}
    \frac{i}{2MT}\sum_{kl}r^kp^{l+1}c_{1kl}&=0,\\
    \sum_{kl}r^{k+1}p^lc_{2kl}&=0.
\end{align}
From these we can conclude that $c_{1kl}=0$ and $c_{2kl}=0$. Notably this also holds for $k,l=0$, which means the off-diagonal elements of $O$ vanish.

From Eq.~\eqref{eq:c13} we can show that $c_{3kl}=0$ if $k,l\neq 0$. To establish $c_{0kl}$ we use the conditions for $C^{0(\dagger)}$.
Writing the commutators and the matrix multiplications, we obtain four equations for $c_{0kl}$ and $c_{3kl}$:
\begin{align}
    \sum_{kl}\left[\left(r+\frac{ip}{2MT}+\frac{\Delta V_{so}}{4T}r\right)r^kp^lc_{0kl}-r^kp^l\left(r+\frac{ip}{2MT}+\frac{\Delta V_{so}}{4T}r\right)c_{3kl}\right]=0,\label{eq:c01}\\
    \sum_{kl}\left[\left(r+\frac{ip}{2MT}+\frac{\Delta V_{os}}{4T}r\right)r^kp^lc_{3kl} - r^kp^l\left(r+\frac{ip}{2MT}+\frac{\Delta V_{os}}{4T}r\right)c_{0kl}\right]=0,\label{eq:c02}\\
    \sum_{kl}\left[\left(r-\frac{ip}{2MT}+\frac{\Delta V_{os}}{4T}x\right)r^kp^lc_{0kl}-r^kp^l\left(r-\frac{ip}{2MT}+\frac{\Delta V_{os}}{4T}r\right)c_{3kl}\right]=0,\label{eq:c03}\\
    \sum_{kl}\left[\left(r-\frac{ip}{2MT}+\frac{\Delta V_{so}}{4T}x\right)r^kp^lc_{3kl}-r^kp^l\left(r-\frac{ip}{2MT}+\frac{\Delta V_{so}}{4T}r\right)c_{0kl}\right]=0.\label{eq:c04}
\end{align}
We can use that we already know that all $c_{3kl}$ except $c_{300}$ vanish. Adding together 
\eqref{eq:c01} and \eqref{eq:c04}, and \eqref{eq:c02} and \eqref{eq:c03} respectively to obtain the two equations
\begin{align}
    \sum_{kl}\left[\left[r+\frac{\Delta V_{so}}{4T}r,r^kp^l\right]c_{0kl}+\frac{i}{2MT}\{p,r^kp^l\}c_{0kl} - \frac{ip}{MT}c_{300}\right]=0,\\
    \sum_{kl}\left[\left[r-\frac{\Delta V_{so}}{4T}r,r^kp^l\right]c_{0kl}-\frac{i}{2MT}\{p,r^kp^l\}c_{0kl} + \frac{ip}{MT}c_{300}\right]=0,
\end{align}
where we have used that $\Delta V_{so}=-\Delta V_{os}$. Adding together also these two equations, we arrive at 
\begin{equation}
    \sum_{kl}[r,r^kp^l]c_{0kl}=\sum_{kl}ilr^kp^{l-1}c_{0kl}=0,
\end{equation}
from which it follows that $c_{0kl}=0$ for $l\neq 0$. To constrain the coefficients for $l=0$, we subtract \eqref{eq:c01} from \eqref{eq:c04}, and \eqref{eq:c02} from \eqref{eq:c03} respectively we obtain
\begin{align}
    \sum_{kl}\left[\frac{i}{2MT}[p,r^kp^l] + \left\{r+\frac{\Delta V_{so}}{4T}r,r^kp^l\right\}\right]c_{0kl} - 2\left(r+\frac{\Delta V_ {so}}{4T}r\right)c_{300} = 0,\\
    \sum_{kl}\left[\frac{i}{2MT}[p,r^kp^l]-\left\{r+\frac{\Delta V_{os}}{4T}r, r^kp^l\right\}\right]c_{0kl} + 2\left(r+\frac{\Delta V_{os}}{4T}r\right)c_ {300}=0.
\end{align}
If we now subtract these two equations and use $\Delta V_{so} = - \Delta V_{os}$ we get
\begin{equation}
    \sum_{kl}2\{r,r^kp^l\}c_{0kl} -4rc_{300} = 0.
\end{equation}
We can simplifly the anticommutator using $[p^l,r]=-ilp^{l-1}$ giving
\begin{equation}
    \sum_{kl}\left(2r^{k+1}p^l-ilr^kp^{l-1}\right)c_{0kl} - 2rc_{300}=0.
\end{equation}
Since for $l\neq 0$ the $c_{0kl}$ vanish, we can set $l=0$. In this case, we get
\begin{equation}
    \sum_{k}r^{k+1}c_{0k0}=rc_{300},
\end{equation}
which leads to $c_{0k0}=0$ for $k\neq 0$ and $c_{00}=c_{300}$. Since the only non-zero coefficients are $c_{000}=c_{300}$, it follows that $O$ is proportional to $\mathds{1}$.

\paragraph{pNRQCD at LO in $E/(\pi T)$.}
In the case of the pNRQCD equation at LO in $E/(\pi T)$, we proceed similarly to the NLO case. We consider the one dimensional case and write $O$ like in eq.~\eqref{eq:genopqcd}.
Explicitly calculating $[C^1,O]=0$ leads in this case to three equations including
\begin{align}
    &\sum_{kl}c_{2kl}r^{k+1}p^l=0,\\
    &\sum_{kl}c_{3kl}\left(r^{k+1}p^l - r^kp^lr\right)=\sum_{kl}c_{3kl}ilr^kp^{l-1}=0,
\end{align}
where for the last equality, we commute $r$ to the left. From this we conclude $c_{2kl}=0$ and $c_{3kl}=0$ for $l\neq 0$. Considering the condition $[C^0,O]=0$ with $c_{2kl}=0$ leads to 
\begin{align}
    &\sum_{kl}c_{1kl}r^{k+1}p^l=0,\label{eq:c1klLOqcd}\\
    &\sum_{kl}c_{0kl}r^{k+1}p^l-c_{3kl}r^kp^lr=0.\label{eq:c0klLOqcd}
\end{align}
From \eqref{eq:c1klLOqcd}, we conclude that $c_{1kl}=0$. Since $c_{3kl}=0$ for $l\neq 0$, it follows from \eqref{eq:c0klLOqcd} that also $c_{0kl}=0$ for $l\neq 0$, since otherwise there is a power of $p$ which does not cancel. 
Therefore, we are left with
\begin{equation}
    \sum_{k}(c_{0k0}-c_{3k0})r^{k+1}=0,
\end{equation}
which lets us conclude that $c_{0k0}=c_{3k0}$. At this point, since $O$ can only consist of powers of $r$, we can invoke the commutation relation with the Hamiltonian. This leads to an equation containing $c_{0k0}$ and $c_{3k0}$ and forces both to vanish for $k\neq0$. Since we already showed $c_{000}=c_{300}$, we therefore again find that only $O\propto\mathds{1}$ fulfills the commutation relation.

We note that for the NLO case, the commutation relations with the Lindblad operators were sufficient since, in that case, the Lindblad operators contain both $r$ and $p$ and can, therefore, cover the whole operator space. In the LO case, however, the Lindblad operators only depend on $r$, which requires invoking the commutation relation with the Hamiltonian.

\section{Impact of time discretization}
\label{app:dt}

\begin{figure}
    \centering
\includegraphics[width=0.75\textwidth]{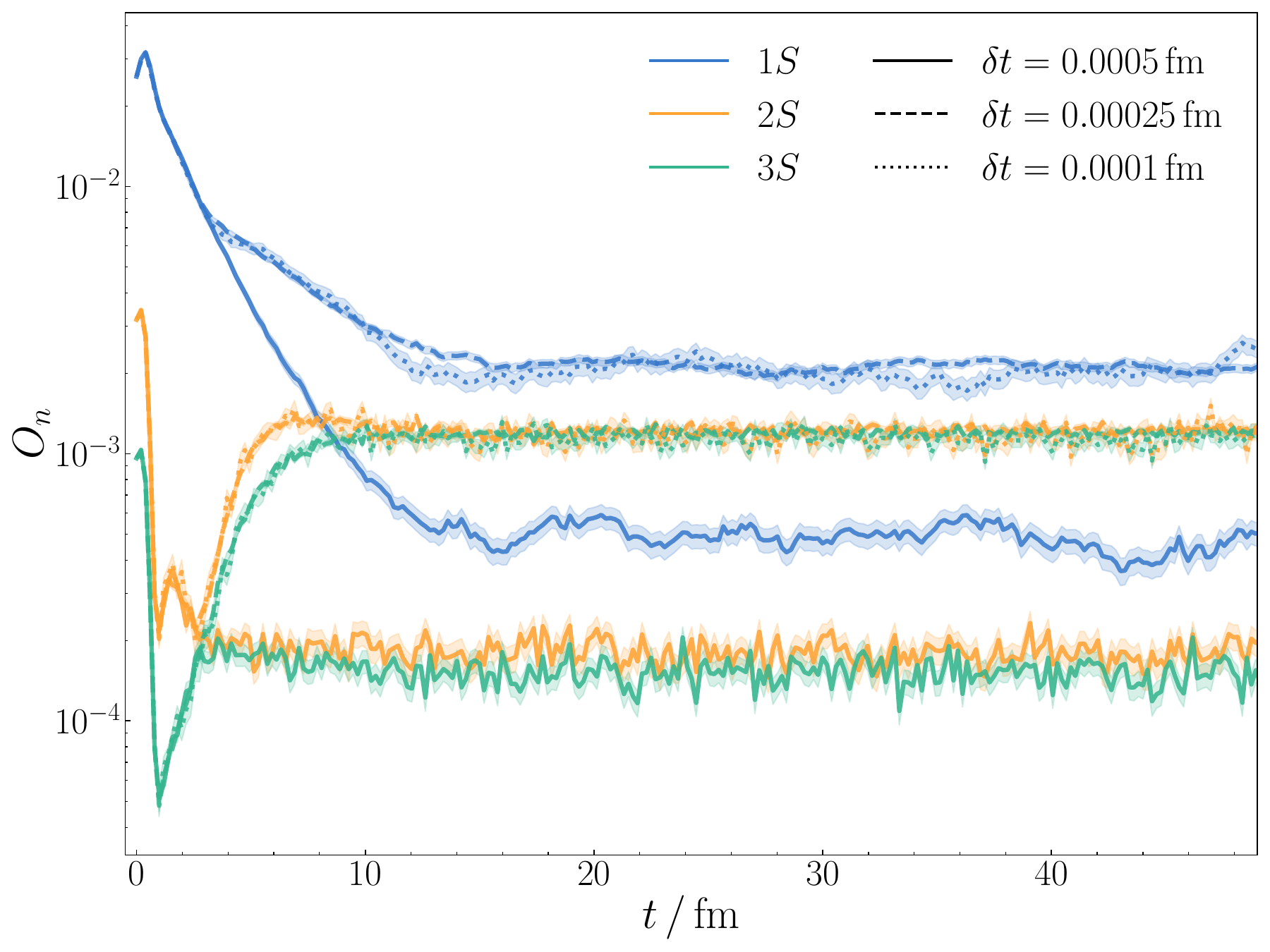}
    \caption{The evolution of the overlaps $O_n$ from the pNRQCD master equation at NLO with a fixed temperature of $T=\SI{450}{\MeV}$. The simulations are performed at $l_\text{max}=3$ and we compare different values of the timestep $\delta t$. We observe that for $\delta t=\SI{0.0005}{\femto\meter}$ the overlaps lie significantly lower than the other timesteps, while $\delta t=\SI{0.00025}{\femto\meter}$ and $\delta t=\SI{0.0001}{\femto\meter}$ agree within statistical uncertainties.}
    \label{fig:A1_QCDNLO_dtcomp}
\end{figure}

As discussed in section~\ref{sec:results}, a small enough time step $\delta t$ is required in order to not underestimate the number of quantum jumps and to achieve convergence of the implicit $H_\text{eff}\delta t$ expansion. Furthermore, since the width $\Gamma = \sum C^{n\dagger}_i C^n_i$ depends on the angular momentum, the correct $\delta t$ will also depend on the angular momentum, which is being evolved. Since transitions between different angular momentum states take place, we have to choose $\delta t$ small enough so that the largest angular momentum value $l_\text{max}$ is still being simulated accurately. This means that we have to choose smaller $\delta t$ for larger $l_\text{max}$. To ensure the choice of a suitable $\delta t$ we perform simulations with decreasing $\delta t$ for the different $l_\text{max}$. In figure~\ref{fig:A1_QCDNLO_dtcomp}, we show the evolution of the overlaps in the case of $l_\text{max}=3$ for three different choices $\delta t = \SI{0.0005}{\femto\meter}, \SI{0.00025}{\femto\meter}, \SI{0.0001}{\femto\meter}$. We observe, that $\delta t=\SI{0.0005}{\femto\meter}$ leads to smaller overlaps, while $\delta t=\SI{0.00025}{\femto\meter}$ and $\delta t=\SI{0.0001}{\femto\meter}$ agree. From this we deduce, that for $\delta t=\SI{0.00025}{\femto\meter}$ the expansion in $\delta t$ converges. 

\begin{figure}
    \centering
\includegraphics[width=0.95\textwidth]{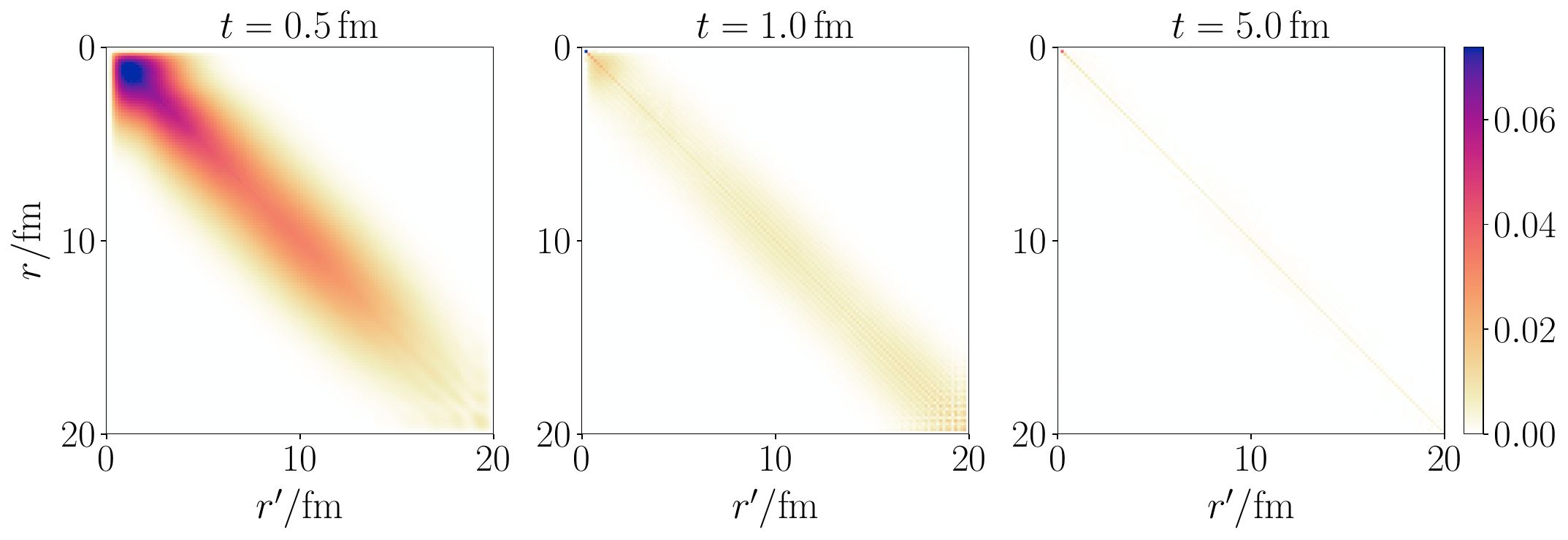}\\
\includegraphics[width=0.95\textwidth]{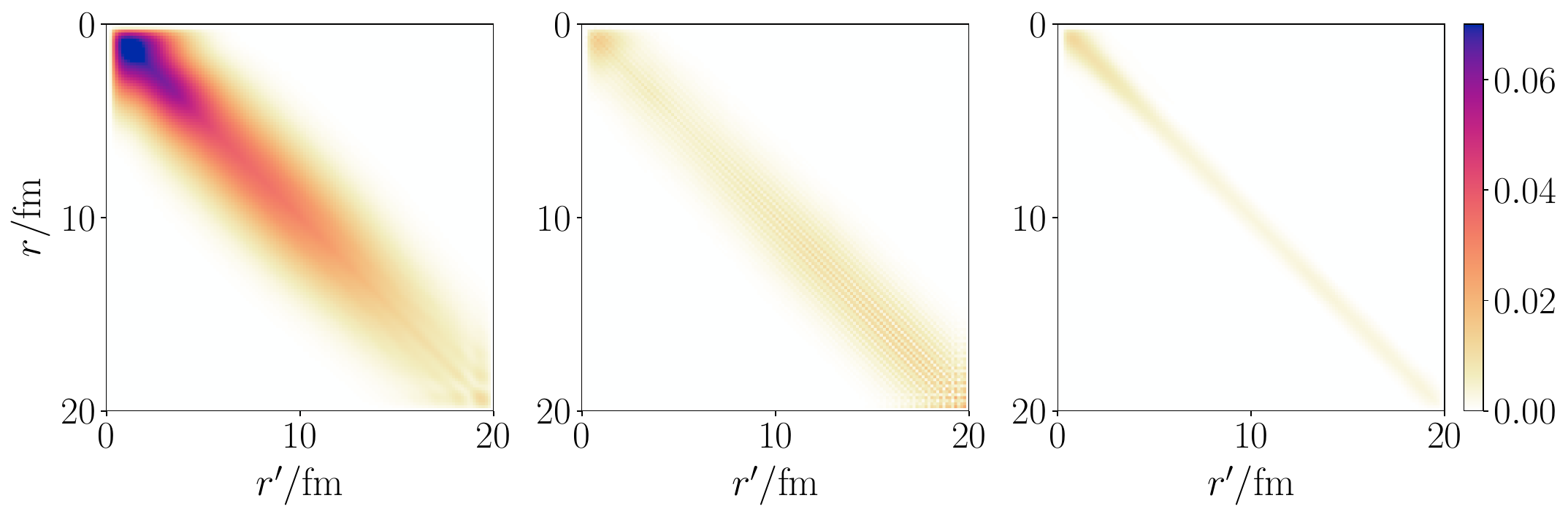}\\
\includegraphics[width=0.95\textwidth]{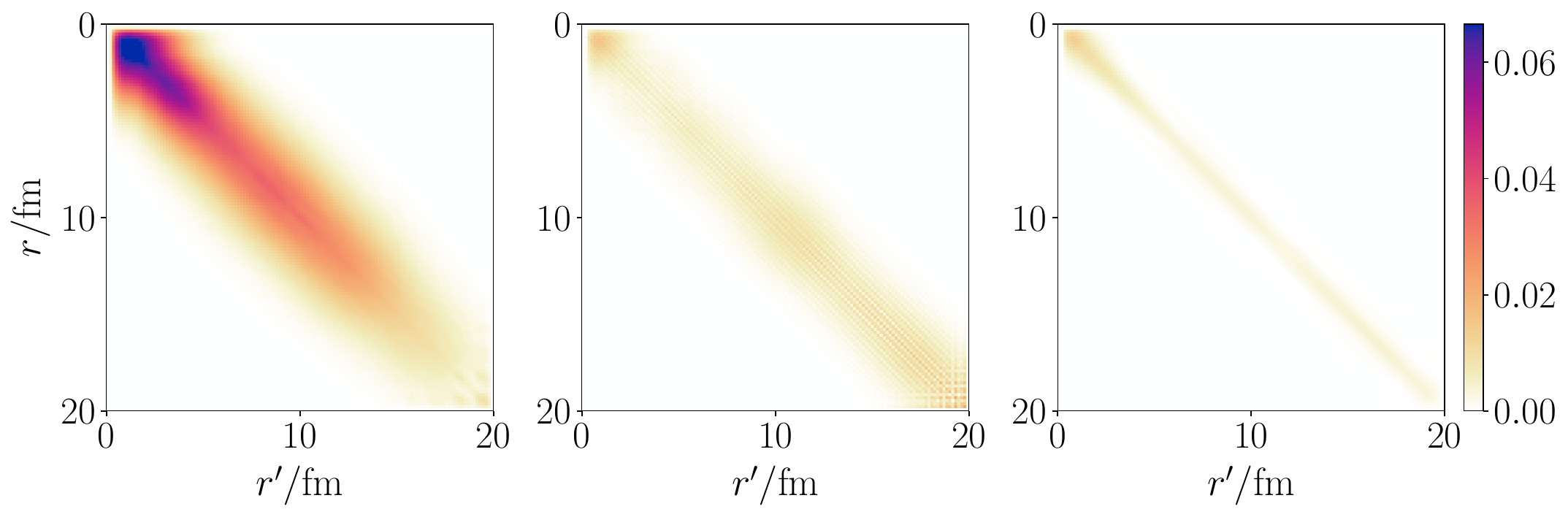}\\
    \caption{The absolute value of the singlet $l=0$ position space density matrix $|\rho^{l=0}_s(r,r^\prime)|$ for simulations of the pNRQCD NLO master equation at $T=\SI{450}{\MeV}$. All simulations are performed at $l_\text{max}=3$ and we show the density matrix at different times, \textit{left} to \textit{right} \SI{0.5}{\femto\meter}, \SI{1}{\femto\meter}, \SI{5}{\femto\meter}. We compare different values for the timestep $\delta t$: \textit{top} to \textit{bottom} $\delta t=\SI{0.0005}{\femto\meter},\SI{0.00025}{\femto\meter},\SI{0.0001}{\femto\meter}$}.
    \label{fig:A2_density}
\end{figure}

To further understand the effect of too large time discretizations, in figure~\ref{fig:A2_density}, we show the evolution of the position space density matrix again with $l_\text{max}=3$ for the different choices of $\delta t$. In the top row, with $\delta t=\SI{0.0005}{\femto\meter}$, we see a strong localization towards $r=0$, characterized by a single pixel corresponding to a large value at the top left origin of the images for $t=\SI{1}{\femto\meter}$ and $t=\SI{5}{\femto\meter}$. Clearly this feature is not physical, but rather a result of the breakdown of the $H_\text{eff}\delta t$ expansion. In the second and third row we show the same position space density matrix for the smaller values of $\delta t$. Here we don't observe the collapse towards $r=0$ but rather a reasonable evolution of the density matrix, which agrees for $\delta t=\SI{0.00025}{\femto\meter}$ and $\delta t=\SI{0.0001}{\femto\meter}$, again supporting the convergence of the $\delta t$ expansion.

Performing a similar analysis for $l_\text{max}=6$ and $l_\text{max}=9$, we find that for these higher $l_\text{max}$ values the smaller timestep of $\delta t=\SI{0.0001}{\femto\meter}$ is required to obtain physical results. All simulations in this study are therefore performed with $\delta t=\SI{0.00025}{\femto\meter}$ for $l_\text{max}=3$ and $\delta t=\SI{0.0001}{\femto\meter}$ for $l_\text{max}=6$ and $l_\text{max}=9$.

These choices for $\delta t$ are smaller than the ones used in previous phenomenological studies with QTraj. This difference is due to the following reasons: Firstly, in phenomenological studies most of the evolution is performed at smaller temperatures, at which the transport coefficient $\kappa=\hat\kappa T^3$ is smaller, which ultimately also leads to a smaller width. Secondly, in phenomenological studies we consider shorter timescales, below $\SI{10}{\femto\meter}$. During these shorter times, the contributions from higher angular momentum states are smaller, since, once higher angular momentum states are exited, it still takes some time, until they contribute considerably to the singlet $l=0$ density.
Finally, in phenomenological simulations we do not impose a tight $l_\text{max}$ cutoff. This reinforces the previous points, since it is more likely to excite higher angular momentum states, without an $l_\text{max}$ cutoff on a timescale of a couple of $\si{\femto\meter}$ only little contributions to the $l=0$ states will come from previously highly excited angular momentum states.

\begin{figure}
    \centering
\includegraphics[width=0.75\textwidth]{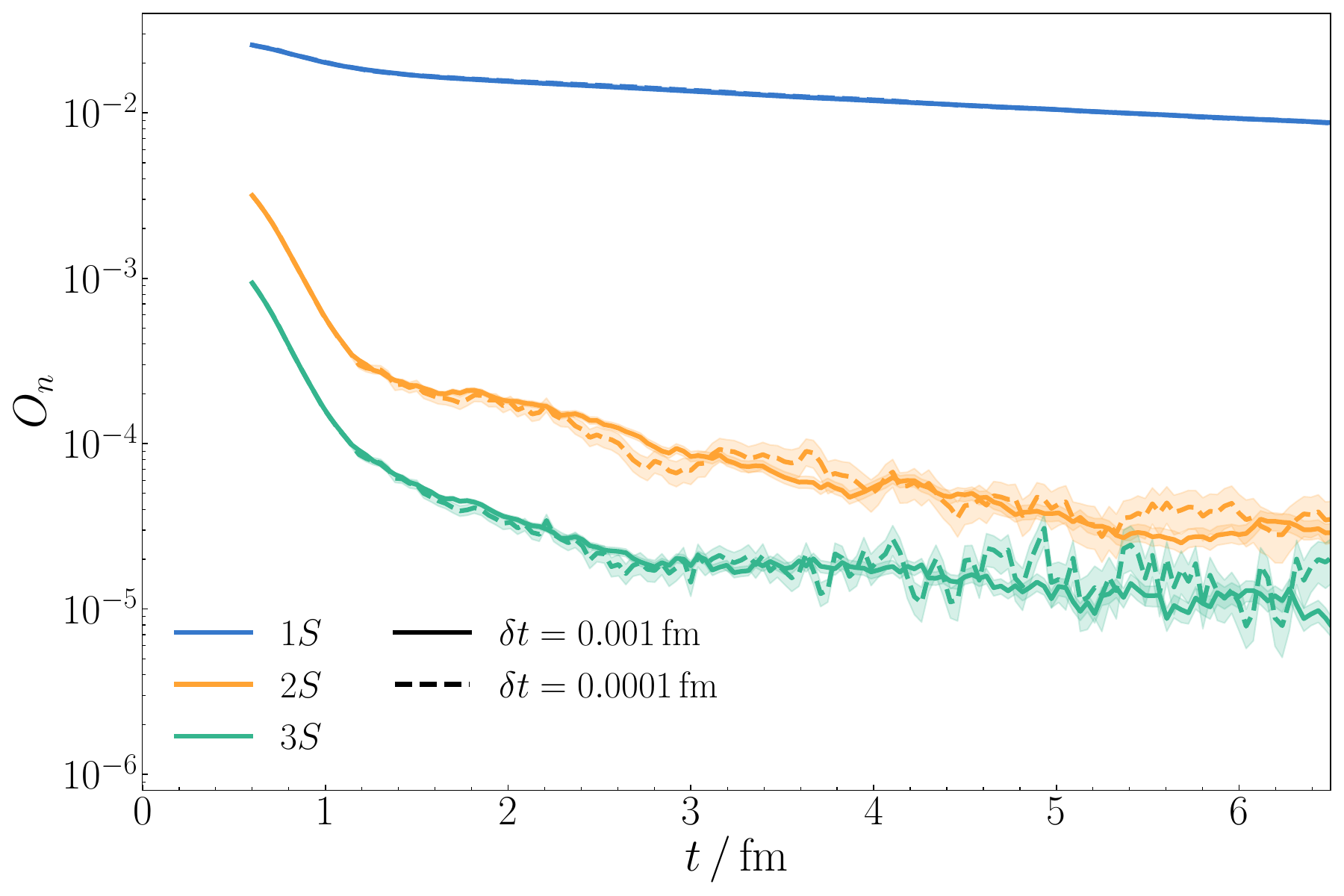}
    \caption{The overlaps $O_n$ from the pNRQCD Lindblad equation at NLO in $E/(\pi T)$ with Bjorken temperature evolution from $T=\SI{425}{\MeV}$ to $\SI{190}{\MeV}$ for the states $nS=1S,2S,3S$. We compare the different time-discretizations $\delta t=\SI{0.001}{\femto\meter}$ and $\delta t=\SI{0.0001}{\femto\meter}$.}
    \label{fig:A3_QCDNLOpheno}
\end{figure}

To show this, in figure~\ref{fig:A3_QCDNLOpheno}, we display the evolution of the $1S,2S$ and $3S$ overlaps in the case of a Bjorken temperature evolution~\cite{Bjorken:1982qr} from $T=\SI{425}{\MeV}$ down to $\SI{190}{\MeV}$. We compare the overlaps for $\delta t=\SI{0.001}{\femto\meter}$ and 
$\delta t=\SI{0.0001}{\femto\meter}$ and see that both results agree within statistical uncertainties, demonstrating that for phenomenological studies the choice of $\delta t=\SI{0.001}{\femto\meter}$ is sufficient.

\section{Lindblad equations in one dimension}
\label{app:1d}

\begin{figure}[h]
    \centering
\includegraphics[width=0.95\linewidth]{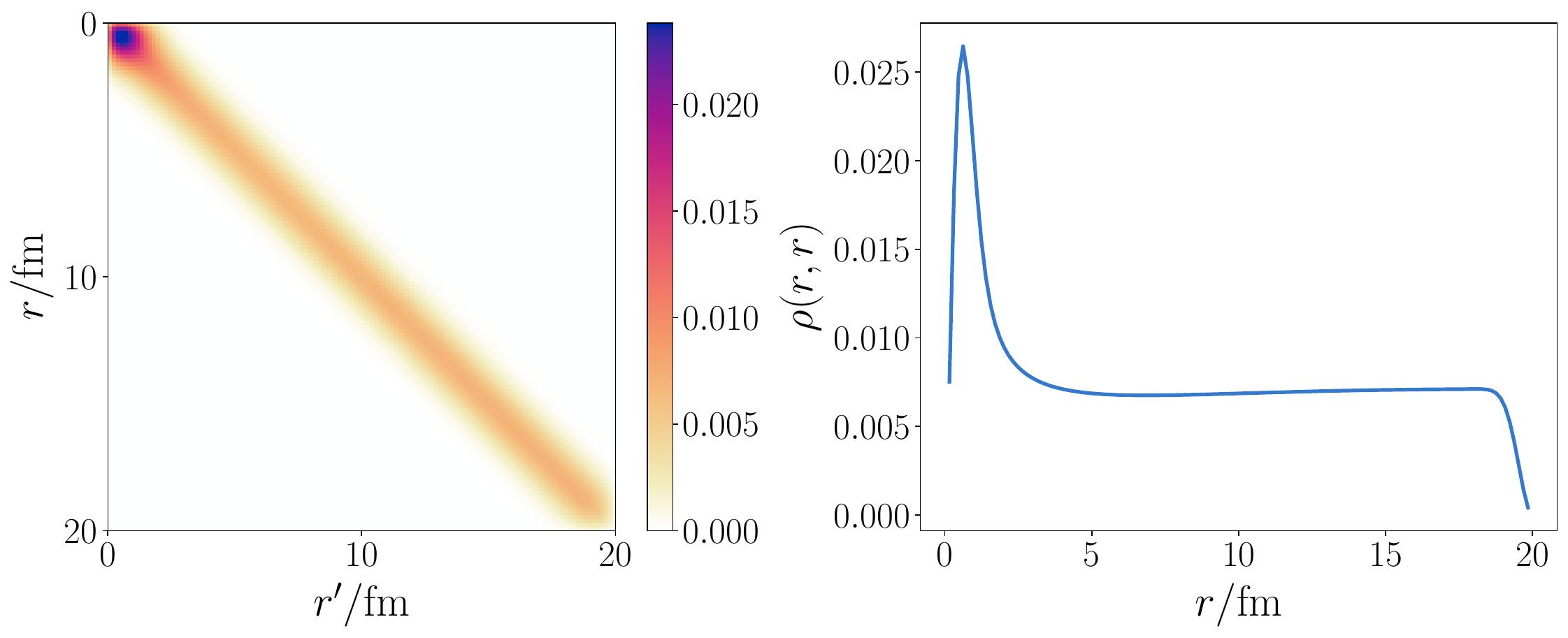}
    \caption{The density matrix of the exact steady state of the one dimensional abelian master equation in a single spatial dimension obtained by finding the eigenvector with a zero eigenvalue of the Liouvillian~\eqref{eq:liouvillian}. \textit{Left:} The full density matrix. \textit{Right:} the diagonal $\rho(r,r)$.}
    \label{fig:diag}
\end{figure}

If we consider the Lindblad equations from section~\ref{sec:mastereqs} in a single dimension, the calculations simplify drastically, since the angular momentum sector does not have to be included. For simplicity, we consider here an "abelianized", or (pNR)QED, version of the above NLO master equation, where we drop the octet sector and consider singlet to singlet transitions. This lead to a Lindblad equation~\eqref{eq:lindblad} with
\begin{align}
    \begin{split}
    H^\text{QED}   &=h_s+\frac{x^2}{2}\gamma + \frac{\kappa}{4MT}\{x,p\},\\
    C^\text{QED}_i &= \sqrt{\kappa}\left(x+\frac{ip}{2MT}\right),
    \end{split}
    \label{eq:QED}
\end{align}
where we consider the same singlet vacuum hamiltonian $h_s$ as in the QCD master equation.
Notably, this master equation is very similar to the Caldeira--Legett model~\cite{CALDEIRA1983374}. While, in principle, the Caldeira--Legett model does not lead to a master equation in Lindblad form, in the limit of large temperatures, one can add a term suppressed by $1/T$ to the equation, which cancels the negative contribution. In this case, it is possible to bring the master equation into Lindblad    form~\cite{Breuer:2002pc}. Up to prefactors this Lindbladian Caldeira--Legget master equation takes the same form as eq.~\eqref{eq:QED}, namely the collapse operators are $C\propto Ax+iBp$ with constant factors $A$ and $B$. Under the assumption that the potential $V(x)$ fulfills  
\begin{equation}
    V(x)-V(x^\prime)\approx V^\prime\left(\frac{x+x^\prime}{2}\right)(x-x^\prime),
    \label{eq:approx1D}
\end{equation}
it is possible to derive an approximate analytic form of the steady state in position space~\cite{Breuer:2002pc}. For a Coulombic potential $V\propto 1/x$ eq.~\eqref{eq:approx1D} is certainly fulfilled on the diagonal $x=x^\prime$ and it also approximately holds for small shifts around the diagonal $x^\prime = x +dx$. Since the steady state is expected to have a band diagonal structure, the approximation is therefore justified for a Coulombic potential. Note, however, that for $x\to 0$, due to the Coulomb singularity, the approximation breaks down. Under the assumption~\eqref{eq:approx1D} the steady state in position space for the master equation~\eqref{eq:QED} can then be written as~\cite{Breuer:2002pc}
\begin{equation}
    \rho_{ss}(x,x^\prime)=N\exp\left[-2\frac{V((x+x^\prime)/2)}{T}-\frac{MT(x-x^\prime)^2}{4}\right].
\end{equation}
While an approximation, this expression qualitatively captures the steady state fairly well. We have large contributions at small $x$ due to the Coulomb potential and a decay for larger $x$, which however becomes very flat due to the $1/x$ in the exponential. Furthermore, we have a decay of off-diagonal elements, with the characteristic width of the band around the diagonal given by $|x-x^\prime|\sim 2/\sqrt{MT}$, as previously noted in ref.~\cite{Delorme:2024rdo}. 
Furthermore, for this one dimensional system, for small lattice sizes, it is possible to obtain the exact steady state, by calculating the eigenvector with zero eigenvalue of the Lindbladian~\eqref{eq:liouvillian}. In figure~\ref{fig:diag}, we exemplarily show the steady state obtained using this method by discretizing the position space on $N=128$ points for a temperature of $T=\SI{450}{\MeV}$. We observe that the steady state takes the characteristic form of a maximum at small $r$, due to the attractive Coulomb potential, and a nearly flat extension until the boundary of the box. This again agrees well with the findings for the one dimensional master equation in ref.~\cite{Delorme:2024rdo}. It furthermore highlights the profound impact the angular momentum sector has in the three-dimensional case, as the centrifugal barrier leads to a maximum above $\SI{10}{\femto\meter}$, see figure~\ref{fig:4_lmaxposition}.

\bibliographystyle{JHEP}
\bibliography{biblio.bib}

\providecommand{\href}[2]{#2}\begingroup\raggedright\begin{thebibliography}{10}

\bibitem{Matsui:1986dk}
T.~Matsui and H.~Satz, \emph{{$J/\psi$ Suppression by Quark-Gluon Plasma Formation}}, \href{https://doi.org/10.1016/0370-2693(86)91404-8}{\emph{Phys. Lett. B} {\bfseries 178} (1986) 416}.

\bibitem{Rothkopf:2019ipj}
A.~Rothkopf, \emph{{Heavy Quarkonium in Extreme Conditions}}, \href{https://doi.org/10.1016/j.physrep.2020.02.006}{\emph{Phys. Rept.} {\bfseries 858} (2020) 1} [\href{https://arxiv.org/abs/1912.02253}{{\ttfamily 1912.02253}}].

\bibitem{Akamatsu:2020ypb}
Y.~Akamatsu, \emph{{Quarkonium in quark\textendash{}gluon plasma: Open quantum system approaches re-examined}}, \href{https://doi.org/10.1016/j.ppnp.2021.103932}{\emph{Prog. Part. Nucl. Phys.} {\bfseries 123} (2022) 103932} [\href{https://arxiv.org/abs/2009.10559}{{\ttfamily 2009.10559}}].

\bibitem{Yao:2021lus}
X.~Yao, \emph{{Open quantum systems for quarkonia}}, \href{https://doi.org/10.1142/S0217751X21300106}{\emph{Int. J. Mod. Phys. A} {\bfseries 36} (2021) 2130010} [\href{https://arxiv.org/abs/2102.01736}{{\ttfamily 2102.01736}}].

\bibitem{Andronic:2024oxz}
A.~Andronic et~al., \emph{{Comparative study of quarkonium transport in hot QCD matter}}, \href{https://doi.org/10.1140/epja/s10050-024-01306-6}{\emph{Eur. Phys. J. A} {\bfseries 60} (2024) 88} [\href{https://arxiv.org/abs/2402.04366}{{\ttfamily 2402.04366}}].

\bibitem{Bodwin:1994jh}
G.T.~Bodwin, E.~Braaten and G.P.~Lepage, \emph{{Rigorous QCD analysis of inclusive annihilation and production of heavy quarkonium}}, \href{https://doi.org/10.1103/PhysRevD.55.5853}{\emph{Phys. Rev. D} {\bfseries 51} (1995) 1125} [\href{https://arxiv.org/abs/hep-ph/9407339}{{\ttfamily hep-ph/9407339}}].

\bibitem{Brambilla:1999xf}
N.~Brambilla, A.~Pineda, J.~Soto and A.~Vairo, \emph{{Potential NRQCD: An Effective theory for heavy quarkonium}}, \href{https://doi.org/10.1016/S0550-3213(99)00693-8}{\emph{Nucl. Phys. B} {\bfseries 566} (2000) 275} [\href{https://arxiv.org/abs/hep-ph/9907240}{{\ttfamily hep-ph/9907240}}].

\bibitem{Brambilla:2004jw}
N.~Brambilla, A.~Pineda, J.~Soto and A.~Vairo, \emph{{Effective Field Theories for Heavy Quarkonium}}, \href{https://doi.org/10.1103/RevModPhys.77.1423}{\emph{Rev. Mod. Phys.} {\bfseries 77} (2005) 1423} [\href{https://arxiv.org/abs/hep-ph/0410047}{{\ttfamily hep-ph/0410047}}].

\bibitem{Miura:2019ssi}
T.~Miura, Y.~Akamatsu, M.~Asakawa and A.~Rothkopf, \emph{{Quantum Brownian motion of a heavy quark pair in the quark-gluon plasma}}, \href{https://doi.org/10.1103/PhysRevD.101.034011}{\emph{Phys. Rev. D} {\bfseries 101} (2020) 034011} [\href{https://arxiv.org/abs/1908.06293}{{\ttfamily 1908.06293}}].

\bibitem{Blaizot:2017ypk}
J.-P.~Blaizot and M.A.~Escobedo, \emph{{Quantum and classical dynamics of heavy quarks in a quark-gluon plasma}}, \href{https://doi.org/10.1007/JHEP06(2018)034}{\emph{JHEP} {\bfseries 06} (2018) 034} [\href{https://arxiv.org/abs/1711.10812}{{\ttfamily 1711.10812}}].

\bibitem{Sharma:2019xum}
R.~Sharma and A.~Tiwari, \emph{{Quantum evolution of quarkonia with correlated and uncorrelated noise}}, \href{https://doi.org/10.1103/PhysRevD.101.074004}{\emph{Phys. Rev. D} {\bfseries 101} (2020) 074004} [\href{https://arxiv.org/abs/1912.07036}{{\ttfamily 1912.07036}}].

\bibitem{Brambilla:2023hkw}
N.~Brambilla, M.A.~Escobedo, A.~Islam, M.~Strickland, A.~Tiwari, A.~Vairo et~al., \emph{{Regeneration of bottomonia in an open quantum systems approach}}, \href{https://doi.org/10.1103/PhysRevD.108.L011502}{\emph{Phys. Rev. D} {\bfseries 108} (2023) L011502} [\href{https://arxiv.org/abs/2302.11826}{{\ttfamily 2302.11826}}].

\bibitem{Brambilla:2024tqg}
N.~Brambilla, T.~Magorsch, M.~Strickland, A.~Vairo and P.~Vander~Griend, \emph{{Bottomonium suppression from the three-loop QCD potential}}, \href{https://doi.org/10.1103/PhysRevD.109.114016}{\emph{Phys. Rev. D} {\bfseries 109} (2024) 114016} [\href{https://arxiv.org/abs/2403.15545}{{\ttfamily 2403.15545}}].

\bibitem{Yao:2018nmy}
X.~Yao and T.~Mehen, \emph{{Quarkonium in-medium transport equation derived from first principles}}, \href{https://doi.org/10.1103/PhysRevD.99.096028}{\emph{Phys. Rev. D} {\bfseries 99} (2019) 096028} [\href{https://arxiv.org/abs/1811.07027}{{\ttfamily 1811.07027}}].

\bibitem{Yao:2020eqy}
X.~Yao and T.~Mehen, \emph{{Quarkonium Semiclassical Transport in Quark-Gluon Plasma: Factorization and Quantum Correction}}, \href{https://doi.org/10.1007/JHEP02(2021)062}{\emph{JHEP} {\bfseries 02} (2021) 062} [\href{https://arxiv.org/abs/2009.02408}{{\ttfamily 2009.02408}}].

\bibitem{Greco:2003vf}
V.~Greco, C.M.~Ko and R.~Rapp, \emph{{Quark coalescence for charmed mesons in ultrarelativistic heavy ion collisions}}, \href{https://doi.org/10.1016/j.physletb.2004.06.064}{\emph{Phys. Lett. B} {\bfseries 595} (2004) 202} [\href{https://arxiv.org/abs/nucl-th/0312100}{{\ttfamily nucl-th/0312100}}].

\bibitem{ALICE:2017pbx}
{\scshape ALICE} collaboration, \emph{{$D$-meson azimuthal anisotropy in midcentral Pb-Pb collisions at $\mathbf{\sqrt{s_{\rm NN}}=5.02}$ TeV}}, \href{https://doi.org/10.1103/PhysRevLett.120.102301}{\emph{Phys. Rev. Lett.} {\bfseries 120} (2018) 102301} [\href{https://arxiv.org/abs/1707.01005}{{\ttfamily 1707.01005}}].

\bibitem{CMS:2017vhp}
{\scshape CMS} collaboration, \emph{{Measurement of prompt $D^0$ meson azimuthal anisotropy in Pb-Pb collisions at $\sqrt{{s}_{NN}}$ = 5.02 TeV}}, \href{https://doi.org/10.1103/PhysRevLett.120.202301}{\emph{Phys. Rev. Lett.} {\bfseries 120} (2018) 202301} [\href{https://arxiv.org/abs/1708.03497}{{\ttfamily 1708.03497}}].

\bibitem{ALICE:2019pox}
{\scshape ALICE} collaboration, \emph{{Measurement of $\Upsilon(1{\rm S})$ elliptic flow at forward rapidity in Pb-Pb collisions at $\sqrt{s_{\rm{NN}}}=5.02$ TeV}}, \href{https://doi.org/10.1103/PhysRevLett.123.192301}{\emph{Phys. Rev. Lett.} {\bfseries 123} (2019) 192301} [\href{https://arxiv.org/abs/1907.03169}{{\ttfamily 1907.03169}}].

\bibitem{deJong:2021wsd}
W.A.~de~Jong, K.~Lee, J.~Mulligan, M.~P\l{}osko\'n, F.~Ringer and X.~Yao, \emph{{Quantum simulation of nonequilibrium dynamics and thermalization in the Schwinger model}}, \href{https://doi.org/10.1103/PhysRevD.106.054508}{\emph{Phys. Rev. D} {\bfseries 106} (2022) 054508} [\href{https://arxiv.org/abs/2106.08394}{{\ttfamily 2106.08394}}].

\bibitem{Lee:2023urk}
K.~Lee, J.~Mulligan, F.~Ringer and X.~Yao, \emph{{Liouvillian dynamics of the open Schwinger model: String breaking and kinetic dissipation in a thermal medium}}, \href{https://doi.org/10.1103/PhysRevD.108.094518}{\emph{Phys. Rev. D} {\bfseries 108} (2023) 094518} [\href{https://arxiv.org/abs/2308.03878}{{\ttfamily 2308.03878}}].

\bibitem{Lin:2024eiz}
J.~Lin, D.~Luo, X.~Yao and P.E.~Shanahan, \emph{{Real-time dynamics of the Schwinger model as an open quantum system with Neural Density Operators}}, \href{https://doi.org/10.1007/JHEP06(2024)211}{\emph{JHEP} {\bfseries 06} (2024) 211} [\href{https://arxiv.org/abs/2402.06607}{{\ttfamily 2402.06607}}].

\bibitem{Angelides:2025hjt}
T.~Angelides, Y.~Guo, K.~Jansen, S.~K\"uhn and G.~Magnifico, \emph{{Meson thermalization with a hot medium in the open Schwinger model}},  \href{https://arxiv.org/abs/2501.13675}{{\ttfamily 2501.13675}}.

\bibitem{Miura:2022arv}
T.~Miura, Y.~Akamatsu, M.~Asakawa and Y.~Kaida, \emph{{Simulation of Lindblad equations for quarkonium in the quark-gluon plasma}}, \href{https://doi.org/10.1103/PhysRevD.106.074001}{\emph{Phys. Rev. D} {\bfseries 106} (2022) 074001} [\href{https://arxiv.org/abs/2205.15551}{{\ttfamily 2205.15551}}].

\bibitem{Delorme:2024rdo}
S.~Delorme, R.~Katz, T.~Gousset, P.B.~Gossiaux and J.-P.~Blaizot, \emph{{Quarkonium dynamics in the quantum Brownian regime with non-abelian quantum master equations}}, \href{https://doi.org/10.1007/JHEP06(2024)060}{\emph{JHEP} {\bfseries 06} (2024) 060} [\href{https://arxiv.org/abs/2402.04488}{{\ttfamily 2402.04488}}].

\bibitem{Brambilla:2016wgg}
N.~Brambilla, M.A.~Escobedo, J.~Soto and A.~Vairo, \emph{{Quarkonium suppression in heavy-ion collisions: an open quantum system approach}}, \href{https://doi.org/10.1103/PhysRevD.96.034021}{\emph{Phys. Rev. D} {\bfseries 96} (2017) 034021} [\href{https://arxiv.org/abs/1612.07248}{{\ttfamily 1612.07248}}].

\bibitem{Brambilla:2017zei}
N.~Brambilla, M.A.~Escobedo, J.~Soto and A.~Vairo, \emph{{Heavy quarkonium suppression in a fireball}}, \href{https://doi.org/10.1103/PhysRevD.97.074009}{\emph{Phys. Rev. D} {\bfseries 97} (2018) 074009} [\href{https://arxiv.org/abs/1711.04515}{{\ttfamily 1711.04515}}].

\bibitem{Brambilla:2022ynh}
N.~Brambilla, M.A.~Escobedo, A.~Islam, M.~Strickland, A.~Tiwari, A.~Vairo et~al., \emph{{Heavy quarkonium dynamics at next-to-leading order in the binding energy over temperature}}, \href{https://doi.org/10.1007/JHEP08(2022)303}{\emph{JHEP} {\bfseries 08} (2022) 303} [\href{https://arxiv.org/abs/2205.10289}{{\ttfamily 2205.10289}}].

\bibitem{Lindblad:1975ef}
G.~Lindblad, \emph{{On the Generators of Quantum Dynamical Semigroups}}, \href{https://doi.org/10.1007/BF01608499}{\emph{Commun. Math. Phys.} {\bfseries 48} (1976) 119}.

\bibitem{Gorini:1975nb}
V.~Gorini, A.~Kossakowski and E.C.G.~Sudarshan, \emph{{Completely Positive Dynamical Semigroups of N Level Systems}}, \href{https://doi.org/10.1063/1.522979}{\emph{J. Math. Phys.} {\bfseries 17} (1976) 821}.

\bibitem{Scheihing-Hitschfeld:2023tuz}
B.~Scheihing-Hitschfeld and X.~Yao, \emph{{Real time quarkonium transport coefficients in open quantum systems from Euclidean QCD}}, \href{https://doi.org/10.1103/PhysRevD.108.054024}{\emph{Phys. Rev. D} {\bfseries 108} (2023) 054024} [\href{https://arxiv.org/abs/2306.13127}{{\ttfamily 2306.13127}}].

\bibitem{Brambilla:2024quarkonium}
N.~Brambilla, M.A.~Escobedo, A.~Islam, M.~Strickland, A.~Vairo and P.V.~Griend, \emph{Anatomy of quarkonium transport coefficients}, {\emph{\textnormal{in preparation, \textsc{TUM-EFT 191/24, FERMILAB-PUB-24-0451-T}}} }.

\bibitem{Choi:1975nug}
M.-D.~Choi, \emph{{Completely positive linear maps on complex matrices}}, \href{https://doi.org/10.1016/0024-3795(75)90075-0}{\emph{Linear Algebra Appl.} {\bfseries 10} (1975) 285}.

\bibitem{Jamiolkowski:1972pzh}
A.~Jamio\l{}kowski, \emph{{Linear transformations which preserve trace and positive semidefiniteness of operators}}, \href{https://doi.org/10.1016/0034-4877(72)90011-0}{\emph{Rept. Math. Phys.} {\bfseries 3} (1972) 275}.

\bibitem{Brody:2013axr}
D.C.~Brody, \emph{{Biorthogonal quantum mechanics}}, \href{https://doi.org/10.1088/1751-8113/47/3/035305}{\emph{J. Phys. A} {\bfseries 47} (2013) 035305} [\href{https://arxiv.org/abs/1308.2609}{{\ttfamily 1308.2609}}].

\bibitem{Baumgartner_2008_I}
B.~Baumgartner, H.~Narnhofer and W.~Thirring, \emph{Analysis of quantum semigroups with gks–lindblad generators: I. simple generators}, \href{https://doi.org/10.1088/1751-8113/41/6/065201}{\emph{Journal of Physics A: Mathematical and Theoretical} {\bfseries 41} (2008) 065201}.

\bibitem{Baumgartner_2008}
B.~Baumgartner and H.~Narnhofer, \emph{Analysis of quantum semigroups with gks–lindblad generators: Ii. general}, \href{https://doi.org/10.1088/1751-8113/41/39/395303}{\emph{Journal of Physics A: Mathematical and Theoretical} {\bfseries 41} (2008) 395303}.

\bibitem{Nigro:2018xov}
D.~Nigro, \emph{{On the uniqueness of the steady-state solution of the Lindblad\textendash{}Gorini\textendash{}Kossakowski\textendash{}Sudarshan equation}}, \href{https://doi.org/10.1088/1742-5468/ab0c1c}{\emph{J. Stat. Mech.} {\bfseries 1904} (2019) 043202} [\href{https://arxiv.org/abs/1803.06279}{{\ttfamily 1803.06279}}].

\bibitem{Spohn_1977}
H.~Spohn, \emph{An algebraic condition for the approach to equilibrium of an open n-level system}, \href{https://doi.org/10.1007/BF00420668}{\emph{Letters in Mathematical Physics} {\bfseries 2} (1977) 33}.

\bibitem{Frigerio:1978gu}
A.~Frigerio, \emph{{Stationary States of Quantum Dynamical Semigroups}}, \href{https://doi.org/10.1007/BF01196936}{\emph{Commun. Math. Phys.} {\bfseries 63} (1978) 269}.

\bibitem{Evans:1977jg}
D.E.~Evans, \emph{{Irreducible Quantum Dynamical Semigroups}}, \href{https://doi.org/10.1007/BF01614091}{\emph{Commun. Math. Phys.} {\bfseries 54} (1977) 293}.

\bibitem{Trushechkin:2021drp}
A.S.~Trushechkin, M.~Merkli, J.D.~Cresser and J.~Anders, \emph{{Open quantum system dynamics and the mean force Gibbs state}}, \href{https://doi.org/10.1116/5.0073853}{\emph{AVS Quantum Sci.} {\bfseries 4} (2022) 012301} [\href{https://arxiv.org/abs/2110.01671}{{\ttfamily 2110.01671}}].

\bibitem{Purkayastha:2024paj}
A.~Purkayastha, G.~Guarnieri, J.~Anders and M.~Merkli, \emph{{On the difference between thermalization in open and isolated quantum systems: a case study}},  \href{https://arxiv.org/abs/2409.11932}{{\ttfamily 2409.11932}}.

\bibitem{Omar:2021kra}
H.B.~Omar, M.A.~Escobedo, A.~Islam, M.~Strickland, S.~Thapa, P.~Vander~Griend et~al., \emph{{QTRAJ 1.0: A Lindblad equation solver for heavy-quarkonium dynamics}}, \href{https://doi.org/10.1016/j.cpc.2021.108266}{\emph{Comput. Phys. Commun.} {\bfseries 273} (2022) 108266} [\href{https://arxiv.org/abs/2107.06147}{{\ttfamily 2107.06147}}].

\bibitem{Molmer:1993ltv}
K.~M\o{}lmer, Y.~Castin and J.~Dalibard, \emph{{Monte Carlo wave-function method in quantum optics}}, \href{https://doi.org/10.1364/JOSAB.10.000524}{\emph{J. Opt. Soc. Am. B} {\bfseries 10} (1993) 524}.

\bibitem{Svetitsky:1987gq}
B.~Svetitsky, \emph{{Diffusion of charmed quarks in the quark-gluon plasma}}, \href{https://doi.org/10.1103/PhysRevD.37.2484}{\emph{Phys. Rev. D} {\bfseries 37} (1988) 2484}.

\bibitem{Moore:2004tg}
G.D.~Moore and D.~Teaney, \emph{{How much do heavy quarks thermalize in a heavy ion collision?}}, \href{https://doi.org/10.1103/PhysRevC.71.064904}{\emph{Phys. Rev. C} {\bfseries 71} (2005) 064904} [\href{https://arxiv.org/abs/hep-ph/0412346}{{\ttfamily hep-ph/0412346}}].

\bibitem{Scardina:2017ipo}
F.~Scardina, S.K.~Das, V.~Minissale, S.~Plumari and V.~Greco, \emph{{Estimating the charm quark diffusion coefficient and thermalization time from D meson spectra at energies available at the BNL Relativistic Heavy Ion Collider and the CERN Large Hadron Collider}}, \href{https://doi.org/10.1103/PhysRevC.96.044905}{\emph{Phys. Rev. C} {\bfseries 96} (2017) 044905} [\href{https://arxiv.org/abs/1707.05452}{{\ttfamily 1707.05452}}].

\bibitem{Bjorken:1982qr}
J.D.~Bjorken, \emph{{Highly Relativistic Nucleus-Nucleus Collisions: The Central Rapidity Region}}, \href{https://doi.org/10.1103/PhysRevD.27.140}{\emph{Phys. Rev. D} {\bfseries 27} (1983) 140}.

\bibitem{CALDEIRA1983374}
A.~Caldeira and A.~Leggett, \emph{Quantum tunnelling in a dissipative system}, \href{https://doi.org/https://doi.org/10.1016/0003-4916(83)90202-6}{\emph{Annals of Physics} {\bfseries 149} (1983) 374}.

\bibitem{Breuer:2002pc}
H.-P.~Breuer and F.~Petruccione, \emph{The Theory of Open Quantum Systems}, Oxford University Press (2007).

\end{thebibliography}\endgroup

\end{document}